\documentclass[prb,aps,twocolumn,preprintnumbers,floats,amsmath,superscriptaddress]{revtex4}

\usepackage{graphicx}
\usepackage{dcolumn}
\usepackage{bm}
\usepackage{amssymb}
\usepackage{bbold}
\usepackage[latin1]{inputenc}

\def \be{\begin{equation}}
\def \ee{\end{equation}}

\begin{document}

\title{Scattering theory of chiral Majorana fermion interferometry}
\author{Jian Li}
\affiliation{D\'epartement de Physique Th\'eorique, Universit\'e de Gen\`eve, CH-1211 Gen\`eve 4, Switzerland}
\author{Genevi\`eve Fleury}
\affiliation{D\'epartement de Physique Th\'eorique, Universit\'e de Gen\`eve, CH-1211 Gen\`eve 4, Switzerland}
\affiliation{Service de Physique de l' \'Etat Condens\'e (CNRS URA 2464),
IRAMIS/SPEC, CEA Saclay, 91191 Gif-sur-Yvette, France}
\author{Markus B\"uttiker}
\affiliation{D\'epartement de Physique Th\'eorique, Universit\'e de Gen\`eve, CH-1211 Gen\`eve 4, Switzerland}
\date{\today}

\begin{abstract}
Using scattering theory, we investigate interferometers composed of chiral
Majorana fermion modes coupled to normal metal leads. We advance an approach
in which also the basis states in the normal leads are written in terms of
Majorana modes. Thus each pair of electron-hole states is associated with a pair of Majorana modes.
Only one lead Majorana mode couples to the intrinsic Majorana mode whereas its partner is completely reflected. Similarly the remaining Majorana modes are completely reflected but in general mix pair-wise. We demonstrate that the charge current can also be expressed in terms of interference between pairs of
Majorana modes. These two basic facts permit a treatment and understanding
of current and noise signatures of chiral Majorana fermion interferometry in
an especially elegant way. As a particular example of applications, in
Fabry-Perot-type interferometers where chiral Majorana modes form loops,
resonances (anti-resonances) from such loops always lead to peaked
(suppressed) Andreev differential conductances, and negative (positive)
cross-correlations that originate purely from two-Majorana-fermion exchange.
These investigations are intimately related to current and noise signatures
of Majorana bound states.
\end{abstract}

\maketitle

\section{Introduction}\label{sec:intro}

Majorana fermions (MFs)  are their own anti-particles. They have been the subject of theories
for more than seven decades without an experimental signature. However recently it has been suggested that these exotic particles can be discovered in condensed matters systems, in particular in topological superconductors (TSCs) \cite{kitaev_unpaired_2001, read_paired_2000, schnyder_classification_2008, schnyder_classification_2009, qi_topological_2011}. TSCs are superconducting systems that admit exceptional boundary states --owing to nontrivial topology associated with bulk spectra-- inside the quasi-particle excitation gaps. These exceptional boundary states are coherent superpositions of electrons (particles) and holes (anti-particles) which, upon proper compositions, represent realizations of Majorana states. The presence of Majorana states in TSCs is robust because of their deep roots in the global properties of bulk states. This leads to renewed efforts to reveal MFs in laboratories.

\begin{figure}
  \centering
  \includegraphics[width=0.32\textwidth]{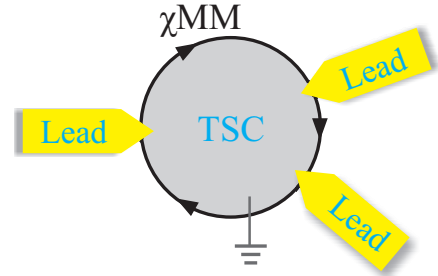}\\
  \caption{Schematic picture of a typical chiral Majorana fermion interferometer composed of a chiral Majorana mode ($\chi$MM) coupled to one or several normal metal leads. The chiral Majorana mode lives at the edge of the underlying topological superconductor (TSC), which has a full pairing gap and is grounded. The interior of the topological superconductor may accommodate a number of vortices, each permitting a magnetic flux quantum to penetrate through and containing a Majorana bound state. Majorana fermions traveling in the chiral Majorana mode pick up a phase that encodes (the parity of) the number of vortices, then scatter into normal leads. To generate charge current and noise Majorana fermions interfere pairwise.}\label{fig:setup}
\end{figure}

Two examples of TSCs are chiral p-wave superconductors in one dimension (1D) and two dimensions (2D). In 1D, a chiral p-wave superconductor may accommodate a pair of Majorana bound states (MBSs) separately at its two ends \cite{kitaev_unpaired_2001}. The energy of this pair of MBSs at large separation is exponentially close to the chemical potential --by definition the zero energy-- of the superconductor, and each MBS hence provides a container for a MF to stay. Recently, systems equivalent to 1D p-wave superconductors have been proposed based on proximity effects in 2D topological insulators \cite{fu_josephson_2009, nilsson_splitting_2008}, or even more close to experimental reality, in 1D semiconducting quantum wires \cite{sau_generic_2010, lutchyn_majorana_2010, oreg_helical_2010}. To detect MBSs in these systems \cite{akhmerov_quantized_2011, wimmer_quantum_2011}, and possibly to manipulate the MFs therein, is of great interest in current research \cite{alicea_non-abelian_2011, flensberg_non-abelian_2011,beenakker_review_2012}.
2D chiral p-wave superconductors \cite{read_paired_2000}, and equivalent systems based on the proximity effect in topological insulators \cite{fu_superconducting_2008} or semiconductors \cite{sau_generic_2010}, are hosts for chiral Majorana modes ($\chi$MMs), which are gapless, charge-neutral edge excitations. $\chi$MMs can serve as coherent transmission channels for MFs, and hence are ideal ingredients for building MF interferometers \cite{fu_probing_2009, akhmerov_electrically_2009, law_majorana_2009, struebi_interferometric_2011} (see, e.g., Fig. \ref{fig:setup}). Unlike conventional (electronic) interferometers, these MF interferometers do not permit an arbitrary magnetic flux insertion due to the presence of the underlying superconductors. Instead, magnetic flux in the MF interferometers is trapped in superconducting vortex cores and remains quantized in integer multiples of a flux quantum. In addition, each vortex here allows for one MBS \cite{kopnin_mutual_1991, read_paired_2000, ivanov_non-abelian_2001}, and the parity of the number of vortices gives rise to a $\mathbb{Z}_2$ -type of MF interferometry \cite{fu_probing_2009, akhmerov_electrically_2009, law_majorana_2009, struebi_interferometric_2011}.

Both examples mentioned above belong to the same universality class, labeled $D$ in the Altland-Zirnbauer classification \cite{altland_nonstandard_1997}. This means that in these systems time-reversal symmetry is broken but particle-hole symmetry (PHS) --as far as the Bogoliubov-de Gennes theory is concerned-- is present. Effectively these systems are spinless, and the PHS operator can be simply expressed as $\Xi = \sigma_x \mathcal{K}$ with $\mathcal{K}$ the complex conjugate operator and $\sigma_x$ the Pauli matrix reversing the electron and hole components. We will limit ourselves to this universality class in this paper, and use the inherent PHS as the only constraint to the scattering theory that will be developed here.

Scattering theory in mesoscopic physics \cite{landauer_electrical_1970, buettiker_four-terminal_1986, buttiker_scattering_1992} has been proven a very useful tool in dealing with electronic coherent transport phenomena. Recent developments have also shown powerful application of scattering theory to the topological classification of condensed matter \cite{fulga_scattering_formula_2011,meidan_topological_2011,fulga_scattering_theory_2011}, extending an approach known from adiabatic pumping processes \cite{graf_topological_2008}. In this paper we employ scattering theory to investigate interferometers built on $\chi$MMs. We pursue especially a general understanding of how MFs --being exotic (quasi-) particles which are charge-neutral-- can be involved in charge transport. For such a purpose we adopt consistently the Majorana basis in our treatment, where the scattering theory of (charge) current and noise takes interestingly an off-diagonal form. Namely, it is the pairwise interference between transmitted MFs, instead of the self-interference of individual particles (as for electrons or holes), that plays the central role in observables like current and noise.

In the following, we will first analyze the scattering at a junction between a $\chi$MM and a normal metal lead, which turns out to be also pairwise in the properly-chosen Majorana basis. In other words, the scattering at the junction can be significantly simplified in terms of both its physical picture and its mathematical parametrization. Next we will apply the simplified scattering picture to compute current and noise in several examples of chiral MF interferometers, including prototypes like Fabry-Perot \cite{law_majorana_2009}, Mach-Zehnder \cite{fu_probing_2009, akhmerov_electrically_2009} and Hanbury Brown-Twiss two-particle  interferometers\cite{struebi_interferometric_2011}. The emphasis of these calculations is placed on demonstrating the way Majorana scattering amplitudes enter the expressions for current and noise, and interpreting the results in a consistent manner. Most of the examples used in this paper have been discussed in the literature, therefore we will not dwell on detailed discussions about the implications of specific results. Nevertheless our approach shows its elegance in dealing with chiral MF interferometers and in understanding them.

\section{Junction between a chiral Majorana mode and a normal lead}
\label{sec:contact}

To address the question how MFs can be involved in charge transport, it is necessary to analyze how they are contacted with a normal (metal) lead which is indispensable for measurements. In the chiral MF interferometry, where the propagating pathways for MFs can be devised to serve various particular interests, junctions between $\chi$MMs and normal leads become a universal, as well as crucial, ingredient and deserve a dedicated investigation. In this section we will focus on such a contact problem, leaving the propagation of chiral MFs to be handled in the next section with specific examples of interferometers.

To this end we consider a $\chi$MM tunneling-coupled to a normal lead which contains $N$ transmission modes around the Fermi energy and is connected to an electron reservoir with chemical potential $\mu$ at its far end \footnote{In the literature there is another type of junctions consisting of one unidirectional ``normal" mode and two Majorana modes with opposite chirality \cite{fu_probing_2009, akhmerov_electrically_2009}. The scattering theory for this type of junctions will be discussed in Sec. \ref{ssec:mz_hbt}.}. The junction is described by the scattering theory adapted to deal with interfaces between normal and superconducting systems \cite{anantram_current_1996}. Namely, the $N$ transmission modes in the normal lead are artificially doubled into their electron and hole copies, which are then treated independently (but constrained by the PHS; see below) with different occupation functions
\begin{align}
\label{eq:fermi}
  f_e(E) &= \frac{1}{e^{\beta(E-\delta\mu)}+1},\\
  f_h(E) &= 1-f_e(-E) = \frac{1}{e^{\beta(E+\delta\mu)}+1},
\end{align}
where $\beta = 1/k_B T$ with $k_B$ the Boltzmann constant and $T$ the temperature, $\delta\mu \equiv \mu - \mu_s$ with $\mu_s$ the chemical potential in the superconductor. Throughout this paper we assume $\delta\mu$ is much smaller than the superconducting gap, $\mu_s$ is fixed by grounding the superconductor, and energy $E$ is always measured relative to $\mu_s$. In order to remove the artificial doubling from actual physical effects, it suffices to count only contributions by states with $E\ge0$.

Specifically, the scattering matrix, $S$, for the junction at energy $E$ is defined by
\be
\label{eq:sm_eh}
  \begin{pmatrix}
    \gamma^{(+)}(E) \\
    \bm{\psi}_{e}^{(+)}(E) \\
    \bm{\psi}_{h}^{(+)}(E) \\
  \end{pmatrix}
  = S(E)
  \begin{pmatrix}
    \gamma^{(-)}(E) \\
    \bm{\psi}_{e}^{(-)}(E) \\
    \bm{\psi}_{h}^{(-)}(E) \\
  \end{pmatrix}
\ee
where $\gamma$ is the annihilation operator for chiral MFs, $\bm{\psi}_{e/h} \equiv ({\psi}_{1e/h},{\psi}_{2e/h},...,{\psi}_{Ne/h})^T$ is the $N$-vector of electron/hole annihilation operators in the normal lead and the subscripts $(+/-)$ always stand for the outgoing/incoming states. The operators here are each normalized by (the square root of) the velocity of the corresponding mode, such that $S$ is a ($2N+1$)-dimensional unitary matrix as a result of quasi-particle probability current conservation. We emphasize that $S$ by itself does not respect charge conservation because of the implicit presence of the TSC upon which the $\chi$MM has to live.

Often it turns out to be more convenient to work with the scattering matrix in an alternative basis --the basis is called the Majorana basis and hence the scattering matrix is denoted by $S_M$-- which is defined as follows
\be
\label{eq:sm_mf}
  \begin{pmatrix}
    \gamma^{(+)}(E) \\
    \bm{\eta}^{(+)}(E) \\
  \end{pmatrix}
  = S_M(E)
  \begin{pmatrix}
    \gamma^{(-)}(E) \\
    \bm{\eta}^{(-)}(E) \\
  \end{pmatrix}
\ee
where
\begin{align}
  \bm{\eta}^{(+/-)}(E)
  \equiv U_N
  \begin{pmatrix}
    \bm{\psi}_{e}^{(+/-)}(E) \\
    \bm{\psi}_{h}^{(+/-)}(E) \\
  \end{pmatrix}, \label{eq:eta}\\
  U_N
  \equiv \frac{1}{\sqrt{2}}
  \begin{pmatrix}
    1 & 1 \\
    i & -i \\
  \end{pmatrix} \otimes \mathbb{1}_N \label{eq:u0}
\end{align}
with $\mathbb{1}_N$ the $N$-dimensional identity matrix acting within the space of electron or hole modes. Evidently $S_M$ is also unitary, and is related to $S$ by a change of basis in the normal lead:
\begin{align}\label{eq:ssm}
  S(E) =
  \begin{pmatrix}
    1 &   \\
      & U_N^\dagger \\
  \end{pmatrix}
  S_M(E)
  \begin{pmatrix}
    1 &   \\
      & U_N \\
  \end{pmatrix}.
\end{align}

The inherent PHS in the current problem imposes that, by definition,
\begin{align}
  \gamma^\dagger(E)&=\gamma(-E), \label{eq:phs_gamma}\\
  \bm{\psi}^{\dagger}_{e}(E) &= \bm{\psi}_{h}(-E), \label{eq:phs_eh}
\end{align}
both of which apply to outgoing and incoming states separately. The latter of the above equations implies
\begin{align}
  \bm{\eta}^\dagger(E) &= \bm{\eta}(-E), \label{eq:phs_eta}
\end{align}
which allows us to regard the scattering modes represented by $\eta$-operators as artificial Majorana modes. The scattering matrices constrained accordingly by the PHS satisfy: in the electron-hole basis,
\begin{align}\label{eq:phs_s}
  \begin{pmatrix}
    1 &   \\
      & \Sigma_x \\
  \end{pmatrix}
  S^{*}(E)
  \begin{pmatrix}
    1 &   \\
      & \Sigma_x \\
  \end{pmatrix}
  = S(-E),
\end{align}
where $\Sigma_x \equiv \sigma_x \otimes \mathbb{1}_N$ with the Pauli matrix $\sigma_x$ interchanges the flavors of electron and hole; in the Majorana basis,
\begin{align}\label{eq:phs_sm}
  S^{*}_M(E)
  = S_M(-E).
\end{align}
It is clear that at $E=0$, $S_M$ becomes a real orthogonal matrix \footnote{Ref. \onlinecite{serban_cre_2010} suggests to call an ensemble of such random matrices CRE (circular real ensemble) to distinguish it from the more familiar CUE (circular unitary ensemble).}.

Indeed, in the rest of this paper, we consider the low-energy limit where the energy dependence of the scattering matrices in Eqs. \eqref{eq:phs_s} and \eqref{eq:phs_sm} can be dropped (accordingly we will also drop the energy dependence of the operators for the scattering modes). As a result $S_M$ is always taken to be a real orthogonal matrix with determinant $1$ (i.e. $S_M \in SO(2N+1)$). This significantly simplifies the theory that will be developed in the following.

\subsection{General strategy}\label{ssec:gen}

In this section our goal is to understand the contact between a $\chi$MM and a normal lead by decomposing the scattering matrix at the junction into its physical part and its (physically) irrelevant part. The physical part is what is relevant to physical observables in the scattering process, and can be taken as the canonical form of the scattering matrix; the irrelevant part by definition does not make physical contributions. To achieve this goal we will work exclusively with scattering matrices in the Majorana basis, $S_M$, in the current section.

The advantage of working in the Majorana basis is that $S_M$, being a real orthogonal matrix in the low-energy limit, is both a faithful and a convenient representation of the scattering processes which involve MFs. Such a representation naturally includes all possible scattering events at the junction without additional constraints as Eq. \eqref{eq:phs_s} (i.e. the set of $S_M$ in the low-energy limit is precisely $SO(2N+1)$, while the corresponding set of $S$ is only a subset of $SU(2N+1)$). This greatly facilitates the decomposition that we will perform in this section. The disadvantage of working in the Majorana basis is that the ``artificial" Majorana modes in the normal lead, defined in Eq. \eqref{eq:eta}, are not associated directly with the distribution of electrons/holes in the reservoir. Therefore we need to return to the electron-hole basis in the next section to compute physical observables such as average current and noise. Nevertheless we will see that the decomposition of $S_M$ pays off not only in simplifying the computations, but also in understanding physically the results.

The idea to decompose a scattering matrix into its physical and irrelevant parts is to notice that the physics of a single scattering event is unchanged under $U(N)$-transformations of the (electron) basis for the $N$ degenerate modes (separately for outgoing and incoming ones) in the normal lead. Decompositions following this idea are conventionally referred to as polar decompositions of a scattering matrix when the physical part contains only the transmission eigenvalues \cite{mello_macroscopic_1988, martin_wave-packet_1992}. In the current problem the same idea gets slightly more complicated by the demand of the PHS, which insists that the valid transformations are of the form
\begin{align}
  \begin{pmatrix}
    \tilde{\bm{\psi}}^{(+)}_{e} \\
    \tilde{\bm{\psi}}^{(+)}_{h} \\
  \end{pmatrix}
  =
  \begin{pmatrix}
    V &   \\
      & V^* \\
  \end{pmatrix}
  \begin{pmatrix}
    \bm{\psi}^{(+)}_{e} \\
    \bm{\psi}^{(+)}_{h} \\
  \end{pmatrix},\label{eq:tf_ehv}\\
  \begin{pmatrix}
    \tilde{\bm{\psi}}^{(-)}_{e} \\
    \tilde{\bm{\psi}}^{(-)}_{h} \\
  \end{pmatrix}
  =
  \begin{pmatrix}
    W &   \\
      & W^* \\
  \end{pmatrix}
  \begin{pmatrix}
    \bm{\psi}^{(-)}_{e} \\
    \bm{\psi}^{(-)}_{h} \\
  \end{pmatrix}\label{eq:tf_ehw}
\end{align}
with $V,W \in U(N)$. In the Majorana basis, if Eq. \eqref{eq:eta} is also imposed on the transformed operators (with tildes), this means
\be
\label{eq:tf_mf}
  \tilde{\bm{\eta}}^{(+)} = \mathcal{V} \bm{\eta}^{(+)},\quad
  \tilde{\bm{\eta}}^{(-)} = \mathcal{W} \bm{\eta}^{(-)},
\ee
where
\begin{align}
\label{eq:vw_mf}
  \mathcal{V}
  =
  U_N
  \begin{pmatrix}
    V &   \\
      & V^* \\
  \end{pmatrix}
  U_N^\dagger,\quad
  \mathcal{W}
  =
  U_N
  \begin{pmatrix}
    W &   \\
      & W^* \\
  \end{pmatrix}
  U_N^\dagger\\
  (\mathcal{V}, \mathcal{W} \in SO(2N)).\nonumber
\end{align}
Accordingly the decomposition of $S_M$ is given by
\begin{align}
\label{eq:sm_dec}
  S_M =
  \begin{pmatrix}
    1 &   \\
      & \mathcal{V}^T \\
  \end{pmatrix}
  \tilde{S}_M
  \begin{pmatrix}
    1 &   \\
      & \mathcal{W} \\
  \end{pmatrix}.
\end{align}
When $\mathcal{V}$ and $\mathcal{W}$ are properly chosen, $\tilde{S}_M$ contains a minimal number of parameters \footnote{Following our strategy, the lower bound of the number of physical parameters is given by $\mbox{Dim}[SO(2N+1)] - 2\,\mbox{Dim}[U(N)] = N$, which turns out to be the actual number of parameters in $\tilde{S}_M$.} that are relevant to physical observables, and hence is the canonical form of $S_M$.

As a consequence of the PHS, the decompositions here do not in general bring $\tilde{S}_M$ into a diagonal form as conventional polar decompositions do (the diagonal form is even impossible in the electron-hole basis because the $\chi$MM cannot be coupled only to a single electron/hole mode at all). Instead the decompositions of $S_M$ turn out to take the form of Euler decompositions: mathematically this means $\tilde{S}_M$ appears to be a combination of uncoupled (2D-)planar rotations; physically this means the scattering of many Majorana modes is essentially pairwise. In the rest of this section, we will derive such an Euler decomposition first in the simplest case when $N=1$, and then in the arbitrary $N$-channel  case.

\subsection{The $N=1$ case}\label{ssec:scdec}

\begin{figure}
  \centering
  \includegraphics[width=0.4\textwidth]{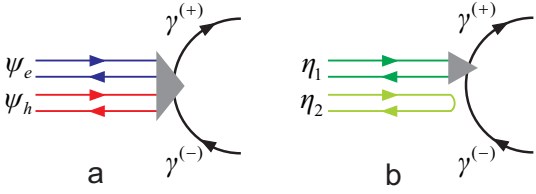}\\
  \caption{Illustrations of the scattering at a junction between a single-mode ($N=1$) normal lead and a $\chi$MM, in the electron-hole basis of the normal lead (a), and in the properly-chosen Majorana basis (b). In the latter basis, only one Majorana mode in the normal lead is coupled to the $\chi$MM, the other is simply reflected with amplitude $1$.}\label{fig:dec1}
\end{figure}

When $N=1$, $S_M$ is of dimension $3$. Namely, $S_M$ is equivalent to a rotation in three dimensions. Such a rotation can be parametrized by three parameters, known as the Euler angles, as follows
\be
\label{eq:euler1}
  S_M =
  \left(\begin{array}{cc}
    1 & 0 \\
    0 & R(\alpha)
  \end{array}\right)
  \left(\begin{array}{cc}
    R(\theta) & 0 \\
    0 & 1
  \end{array}\right)
  \left(\begin{array}{cc}
    1 & 0 \\
    0 & R(\beta)
  \end{array}\right),
\ee
where
\be
\label{eq:r0}
  R(\xi) \equiv
  \left(\begin{array}{cc}
    \cos\xi & -\sin\xi \\
    \sin\xi & \cos\xi
  \end{array}\right)\quad (\xi = \alpha, \beta, \theta)
\ee
is a planar rotation by an angle $\xi$.

The above decomposition is exactly what is dictated by Eq. \eqref{eq:sm_dec}. To see this, simply notice that
\be
\label{eq:r0u0}
  U_1^\dagger R(\xi) U_1 =
  \left(\begin{array}{cc}
    e^{-i\xi} &  \\
     & e^{i\xi}
  \end{array}\right),
\ee
thus we can identify $\mathcal{V} = R(-\alpha)$ and $\mathcal{W} = R(\beta)$. Namely, if we redefine the electron and hole operators by trivial gauge transformations (in the normal lead)
\begin{align}
\label{eq:eh_redef1}
  \begin{pmatrix}
    \tilde{\psi}^{(+)}_{e} \\
    \tilde{\psi}^{(+)}_{h} \\
  \end{pmatrix}
  =
  \begin{pmatrix}
    e^{i\alpha}\psi^{(+)}_{e} \\
    e^{-i\alpha}\psi^{(+)}_{h} \\
  \end{pmatrix},\quad
  \begin{pmatrix}
    \tilde{\psi}^{(-)}_{e} \\
    \tilde{\psi}^{(-)}_{h} \\
  \end{pmatrix}
  =
  \begin{pmatrix}
    e^{-i\beta}\psi^{(-)}_{e} \\
    e^{i\beta}\psi^{(-)}_{h} \\
  \end{pmatrix},
\end{align}
and redefine $\tilde{\bm{\eta}}$ according to Eq. \eqref{eq:eta}, then the Majorana scattering matrix contains one single parameter and takes the form
\be
\label{eq:sm_1}
  \tilde{S}_M =
  \begin{pmatrix}
    R(\theta) & 0 \\
    0 & 1
  \end{pmatrix}.
\ee

Eq. \eqref{eq:sm_1} is the canonical form of the Majorana scattering matrix when $N=1$, which says that, effectively, the $\chi$MM $\gamma$ is coupled to one (artificial) Majorana mode $\tilde{\eta}_1$ in the normal lead, leaving the other (artificial) Majorana mode $\tilde{\eta}_2$ to be completely reflected with amplitude $1$ (see Fig. \ref{fig:dec1}).

\subsection{The arbitrary $N$ case}\label{ssec:mcdec}

\begin{figure}
  \centering
  \includegraphics[width=0.45\textwidth]{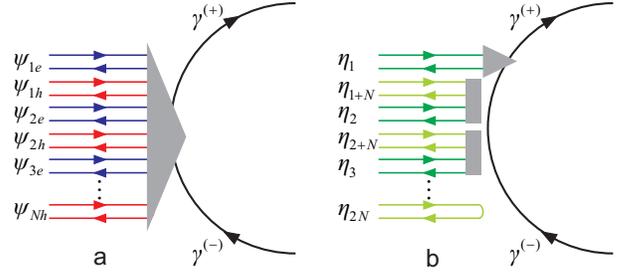}\\
  \caption{Illustrations of the scattering at a junction between a multi-mode ($N>1$) normal lead and a $\chi$MM, in the electron-hole basis for the normal lead (a), and in the properly-chosen Majorana basis (b). In the latter basis, only one Majorana mode in the normal lead is coupled to the $\chi$MM, the others are reflected in pairs and one is reflected with amplitude $1$.}\label{fig:decn}
\end{figure}

When $N>1$, there is no decomposition like Eq. \eqref{eq:euler1} readily available. Nevertheless one can show that (see Appendix \ref{app:dec}), the $(2N+1)$-dimensional Majorana scattering matrix can always be decomposed as Eq. \eqref{eq:sm_dec} with
\be
\label{eq:sm_n}
  \tilde{S}_M = \Bigl[\bigoplus\limits_{i=1}^{N} R(\theta_i)\Bigr] \oplus 1,
\ee
where $R(\theta_1)$ couples the modes $\gamma$ and $\tilde{\eta}_1$, $R(\theta_i)$ with $i>1$ couples the modes $\tilde{\eta}_{i}$ and $\tilde{\eta}_{i+N-1}$, and the last diagonal entry $1$ signifies a complete reflection of mode $\tilde{\eta}_{2N}$. Note that by definition (see Eq. \eqref{eq:eta}), $\tilde{\eta}_i$ and $\tilde{\eta}_{i+N}$ are partners derived from the same normal mode $\tilde{\psi}_{ie/h}$, that is,
\begin{align}
  \tilde{\eta}_i &= \frac{1}{\sqrt{2}}(\tilde{\psi}_{ie} + \tilde{\psi}_{ih}), \label{eq:eta_i}\\
  \tilde{\eta}_{i+N} &= \frac{i}{\sqrt{2}}(\tilde{\psi}_{ie} - \tilde{\psi}_{ih}). \label{eq:eta_in}
\end{align}
The concept of such pairs of Majorana modes turns out to be fundamental in relating Majorana scattering matrices to charge transport, as will be revealed in the next section.

The above canonical form of Majorana scattering matrices should be no surprise after we have understood the $N=1$ case: the $\chi$MM $\gamma$ effectively couples to a single Majorana mode $\tilde{\eta}_1$ -- determined by the specifics of the junction -- in the normal lead, which leaves its partner $\tilde{\eta}_{1+N}$ -- determined solely by $\tilde{\eta}_1$ -- to be coupled to the rest of the Majorana modes in the normal lead; $\tilde{\eta}_{1+N}$ then ``picks" another (single) Majorana mode $\tilde{\eta}_2$ to couple to, leaving in turn $\tilde{\eta}_{2+N}$ behind; this process goes on until the last one $\tilde{\eta}_{2N}$ is left completely alone (see Fig. \ref{fig:decn}). Interestingly, this re-pairing picture is a good analogy to the re-pairing of MFs that happens at the topological phase transition in the Kitaev model for 1D p-wave superconductors \cite{kitaev_unpaired_2001} -- here, the re-pairing of Majorana modes happens when a normal lead makes contact to a 2D p-wave superconductor with a chiral Majorana edge mode.

\section{Current and noise for specific examples}\label{sec:obs}

In this section we compute the average current and the zero-frequency noise power in various prototypical chiral MF interferometers.

To this end we introduce for convenience two matrices: $F$, which is a diagonal matrix containing the Fermi distribution functions for all incoming modes, and $(\Sigma_z)_{\nu}$, which is another diagonal matrix containing the weights for charge current carried by electron and hole modes in lead $\nu$ ($1$ for electron modes and $-1$ for hole modes; $0$ for modes that do not belong to lead $\nu$). Explicitly, we define
\begin{align}
  F &\equiv \bigoplus\limits_{\nu = 1}^{K} F_{\nu}, \label{eq:F}\\
  F_{\nu} &\equiv
  \begin{pmatrix}
    f_{\nu e} & 0\\
    0 & f_{\nu h}
  \end{pmatrix} \otimes \mathbb{1}_{N_{\nu}}, \label{eq:F_nu}\\
  (\Sigma_z)_{\nu} &\equiv
  \bigl(\bigoplus\limits_{\nu^\prime \ne \nu}\mathbb{0}_{\nu^\prime}\bigr)
  \oplus\,\bigl(\sigma_z \otimes \mathbb{1}_{N_{\nu}}\bigr), \label{eq:Sigmaz_nu}
\end{align}
where $K$ is the total number of contacts, $f_{\nu e/h}$ is the Fermi distribution function for electrons/holes in contact $\nu$, $N_{\nu}$ is the number of transmission modes (without double counting) in lead $\nu$, $\mathbb{0}_{\nu^\prime}$ is the empty matrix corresponding to the modes in lead $\nu^\prime$.

We will use the following formulas from the scattering theory adapted for normal-superconducting hybrid systems \cite{buttiker_scattering_1992, anantram_current_1996}: the time-averaged current at normal contact $\alpha$, $\langle \hat{I}_{\alpha} \rangle$, is given by
\begin{align}\label{eq:I}
  I_\alpha = \frac{e}{h}\int_{E\geq 0} dE\,\mathrm{Tr}[F A_\alpha]\;;
\end{align}
the average current flowing through the grounded superconducting contact is given, owing to current conservation, by (omitting the minus sign)
\begin{align}\label{eq:IS}
  I_S = \sum\limits_{\alpha\in\mbox{\scriptsize normal}} I_\alpha\;;
\end{align}
the zero-frequency noise power, defined as $P_{\alpha\beta} = \int_{-\infty}^{\infty}dt\, \frac{1}{2} \langle \delta\hat{I}_\alpha(t)\delta\hat{I}_\beta(0) + \delta\hat{I}_\beta(0)\delta\hat{I}_\alpha(t) \rangle$ with the current fluctuation $\delta\hat{I}_{\nu = \alpha,\beta}(t) = \hat{I}_\nu(t) - \langle \hat{I}_{\nu} \rangle$, is given by
\begin{align}\label{eq:P}
  P_{\alpha\beta} = \frac{e^2}{h}\int_{E\geq 0} dE\,\mathrm{Tr}[F A_\alpha (\mathbb{1}-F) A_\beta]\;.
\end{align}
In the above formulas we have used a shorthand notation $A_{\nu}$ defined as follows
\begin{align}
  A_{\nu} &\equiv (\Sigma_z)_{\nu} - S^{\dagger}(\Sigma_z)_{\nu}\,S. \label{eq:A_nu}
\end{align}
Note that $S$ here is the full scattering matrix in the electron-hole basis, which includes $K$ contacts but does not explicitly include the $\chi$MM (see, e.g., Eq. \eqref{eq:Sl21} in Sec. \ref{ssec:FPI1}). The full Majorana scattering matrix $S_M$ is related to $S$ by
\begin{align}\label{eq:ssmk}
  S = U_{\oplus}^\dagger S_M U_{\oplus},\quad
  U_{\oplus} \equiv \bigoplus\limits_{\nu=1}^K U_{N_\nu}.
\end{align}

Before we proceed presenting the results, we confirm that the transformations in Eqs. \eqref{eq:tf_ehv} and \eqref{eq:tf_ehw} indeed lead to no change in physical observables. That is, we verify that $I_\alpha$ and $P_{\alpha\beta}$ are unchanged upon replacing $S$ by $\tilde{S}$, given by
\begin{eqnarray}
  &S = V_{\oplus}^{\dagger} \tilde{S} W_{\oplus}\:,& \label{eq:stilde}\\
  &V_{\oplus} \equiv
  \bigoplus\limits_{\nu=1}^K
  \begin{pmatrix}
    V_{\nu} &   \\
      & V_{\nu}^* \\
  \end{pmatrix}, \quad
  W_{\oplus} \equiv
  \bigoplus\limits_{\nu=1}^K
  \begin{pmatrix}
    W_{\nu} &   \\
      & W_{\nu}^* \\
  \end{pmatrix}.& \label{eq:VWall}
\end{eqnarray}
The proof is straightforward after noticing that
\begin{align}\label{eq:com}
  [W_{\oplus}, F] = 0\quad \mbox{and} \quad \forall \nu: [V_{\oplus}, (\Sigma_z)_{\nu}] = 0,
\end{align}
where $[,]$ stands for the commutator. Therefore we will adopt only the canonical form $\tilde{S}_M$ in the actual calculations, and also suppress the tildes from now on.

\subsection{A single normal lead probing a chiral Majorana loop: $N=1$}
\label{ssec:l11}

We start with the simplest setup: a single normal lead with a single mode is probing a closed $\chi$MM, or, a chiral Majorana loop; the interior of the TSC -- enclosed by the chiral Majorana loop -- may accommodate $n_v$ vortices, each allowing for one MBS at zero energy. This setup, as well as the setup that will be discussed in the next subsection (Sec. \ref{ssec:FPI1}), has been discussed by Law, Lee and Ng \cite{law_majorana_2009}. Here we reformulate the problem in a consistent Majorana language, with extensions that are intriguing in their own right.

We assume that the MBSs are all deep inside the TSC, such that the coupling between the MBSs and the $\chi$MM or the normal lead, which exponentially decays with respect to their spatial separation, is negligible. With this assumption the vortices are taken into account only as an additional phase $\phi = n_v\pi$ that the chiral MF picks up after moving in a full circle, and the acquired total phase hence is $\varphi = n_v\pi + \pi + 2\pi E/E_l$, where the single $\pi$ comes from the Berry phase, and $E_l = h v_M/L$, with $L$ the circumference of the loop, $v_M$ the group velocity of the $\chi$MM, is the level spacing for the chiral Majorana loop.

\begin{figure}
  \centering
  \includegraphics[width=0.45\textwidth]{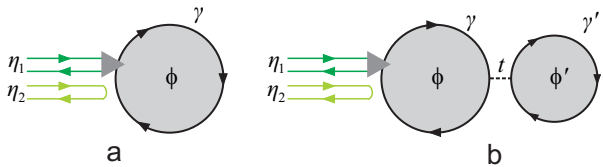}\\
  \caption{Effective scattering for setups with a single-mode ($N=1$) normal lead probing a chiral Majorana loop: the left panel case (a) can be extended into the right panel one (b), to model, e.g., the presence of an additional point contact or scattering at two coupled MBSs (see text for details).}\label{fig:l11}
\end{figure}

Following our discussion in Section \ref{ssec:scdec}, the effective scattering process for the current example in the Majorana language is pictured in Fig. \ref{fig:l11}a, and the Majorana scattering matrix is given by
\begin{equation}
\label{eq:Sl11}
  S_M =
  \begin{pmatrix}
    r_1 & 0 \\
    0 & r_2
  \end{pmatrix},
\end{equation}
where $r_2 = 1$, and
\be
\label{eq:r1}
  r_1=\frac{r_0-e^{i\varphi}}{1-r_0 e^{i\varphi}}
\ee
with $r_0$ the (local) reflection amplitude of $\eta_1$ at the junction connecting to the still unclosed $\gamma$ (namely, $\cos\theta$ in Eq. \eqref{eq:sm_1}). Note that the above Majorana scattering matrix is implicitly energy-dependent because of the energy-dependence of $\varphi$; it is unitary, but not necessarily real except at $E=0$.

Inserting $S_M$ into Eqs. \eqref{eq:I} and \eqref{eq:P}, together with \eqref{eq:ssmk}, we find the average current and the zero-frequency noise power (auto-correlator) to be
\begin{align}
  I &= \frac{e}{h}\int_{E\geq 0} dE\, \bigl[1-\Re(r_1r_2^*)\bigr]\delta{f}(E), \label{eq:Il11}\\
  P &= \frac{e^2}{h}\int_{E\geq 0} dE\, \Bigl\{2\bigl[1-\Re(r_1r_2^*)\bigr]\Theta(E) \nonumber\\
    &\qquad\qquad\qquad\quad + \Im(r_1r_2^*)^2\delta{f}(E)^2\Bigr\}, \label{eq:Pl11}
\end{align}
where
\begin{align}
  \delta{f}(E) &\equiv f_e(E) - f_h(E) = \frac{\sinh{\beta \delta\mu}}{\cosh{\beta E}+\cosh{\beta \delta\mu}}, \label{eq:deltaf}\\
  \Theta(E) &\equiv f_e(E)\bigl[1-f_e(E)\bigr]+f_h(E)\bigl[1-f_h(E)\bigr] \nonumber\\
    &= -k_B T \frac{\partial}{\partial E}\bigl[f_e(E)+f_h(E)\bigr]. \label{eq:Theta}
\end{align}
Note that $\delta{f}(E)$ is positive (negative) definite if $\delta\mu>0$ ($\delta\mu<0$), zero at equilibrium, while $\Theta(E)$ is positive definite at finite temperature. As a convention, we shall intentionally keep the formal complex conjugate of real variables (e.g. $r_2^*$) for a reason that will be clear in a moment.

\subsubsection{Current as interference of Majorana modes}

Let us now examine the expressions for the current Eq. \eqref{eq:Il11} and noise Eq. \eqref{eq:Pl11} closely. The net current in Eq. \eqref{eq:Il11} is evidently attributed to the Andreev process: an incoming electron or hole, weighted effectively by the difference of the Fermi distribution functions $\delta f$, is reflected with ``probability" $\Re(r_1r_2^*) \in [-1,1]$, and the missing charges are absorbed by the superconducting condensate as Cooper pairs. Here, negative $\Re(r_1r_2^*)$ means that a particle with an opposite charge with respect to the incoming particle is reflected with probability $|\Re(r_1r_2^*)|$. The noise in Eq. \eqref{eq:Pl11} has two contributions: the first term in the curly brackets gives the thermal noise which persists at equilibrium at finite temperature; the second term gives the shot noise which vanishes either at equilibrium, where $f_e = f_h$, or upon total normal or Andreev reflections, where $\Im(r_1r_2^*)^2 = 1-\Re(r_1r_2^*)^2 = 0$.

The remarkable expression for the reflection ``probability", $\Re(r_1r_2^*)$, highlights one essential feature of MF transport, namely, charge is always ``carried" by the interference between (a pair of) Majorana modes\footnote{Of course, the real container of charges here is the superconducting condensate, the interference between Majorana modes simply determines how normal leads draw electrons/holes from the container.}, even though an individual Majorana mode is charge-neutral. This can be heuristically understood as follows. Take one (quasi-particle) scattering state given by, in the Majorana basis, $|\Psi(E)\rangle = \frac{1}{\sqrt{2}} \bigl[r_1|\eta_1(E)\rangle + ir_2|\eta_2(E)\rangle\bigr]$; in the electron-hole basis, the same state reads $|\Psi(E)\rangle = a_e|\psi_e(E)\rangle + a_h|\psi_h(E)\rangle$ with $a_e = (r_1+r_2)/2$ and $a_h = (r_1-r_2)/2$; we immediately see that the current density carried by this state is $I(E) = (e/h)(|a_e|^2-|a_h|^2) = (e/h)\Re(r_1r_2^*)$, which is precisely the interference term between the two Majorana modes. When $r_1$ and $r_2$ are identified with the reflection amplitudes in $S_M$, $\Re(r_1r_2^*)$ becomes the reflection ``probability" that appears in Eqs. \eqref{eq:Il11} and \eqref{eq:Pl11}. In order to stress the central importance of interference between Majorana modes in this work, we will on most occasions keep formal expressions similar to $\Re(r_1r_2^*)$ without simplifying expressions written with formally complex conjugate but in reality real variables.

\subsubsection{The zero-temperature low-bias limit}

At zero-temperature and low bias ($\delta\mu = eV \ll E_l$), $\varphi \simeq (n_v+1)\pi$, $r_1 \simeq \pm1$, and Eqs. \eqref{eq:Il11} and \eqref{eq:Pl11} can be reduced to \cite{law_majorana_2009}:
\begin{align}
  &\mbox{for even $n_v$:}\qquad I = P = 0\,; \label{eq:l11_e}\\
  &\mbox{for odd $n_v$:}\qquad I = \frac{2 e^2}{h}\,V,\; P = 0\,. \label{eq:l11_o}
\end{align}
The Fano factor, defined as $F = P/(eI)$, is give by $1+r_1$ and takes the value $2$ if $n_v$ is even, or $0$ if $n_v$ is odd.

Physically the two situations here are realized exactly when the chiral MF self-interferes maximally destructive or constructive upon reflection, hence a stationary state is disallowed or allowed at zero energy. $r_1=\pm1$ therefore corresponds to the reflection of a Majorana mode that is completely off-resonance ($r_1=+1$) or on-resonance ($r_1=-1$) with such a stationary state \cite{law_majorana_2009}. As a consequence, the scattering process is a total normal reflection ($I=0$) or a total Andreev reflection ($I = 2 e^2 V/h$) -- in either case, the shot noise vanishes.

\begin{figure}
  \centering
  \includegraphics[width=0.4\textwidth]{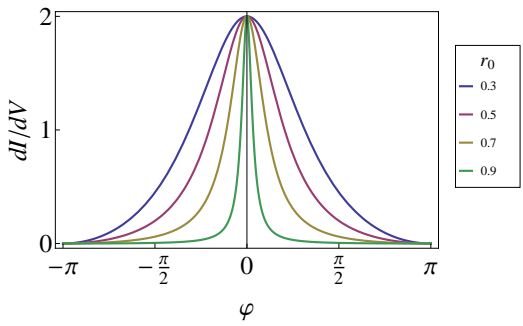}\\
  \caption{Differential conductance, $dI/dV$ in units of $e^2/h$,  as a function of $\varphi = 2\pi eV/E_l + (n_v+1)\pi$ for the setup in Fig. \ref{fig:l11}a. The resonance peak for $\varphi$ equal to an even multiple of $\pi$ and the (anti-resonance) valley for $\varphi$ equal to an odd multiple of $\pi$ is a robust feature regardless of the coupling strength, parametrized by $r_0$, between the normal lead and the $\chi$MM. This plot is essentially a reproduction of the results of Law et al. \cite{law_majorana_2009}. The purpose of this reproduction is to facilitate comparisons with our later results.}\label{fig:Gl11}
\end{figure}

For more general cases beyond the low-bias limit, where $\varphi$ is allowed to vary with energy continuously between an even and an odd multiple of $\pi$, we plot the zero-temperature differential conductance $dI/dV$, as a function of $\varphi = 2\pi eV/E_l + (n_v+1)\pi$ and in units of $e^2/h$, in Fig. \ref{fig:Gl11} with different $r_0$. It is clearly seen that a quantized resonance peak is established at $\varphi = 0$ ($\mbox{mod}\;2\pi$) and completely suppressed at $\varphi = \pi$ ($\mbox{mod}\;2\pi$) \cite{law_majorana_2009}; decreasing $r_0$ broadens the resonance peak but does not remove such a general feature.

\subsubsection{Extended setups}\label{sssec:ext1}

The present simple setup can be slightly extended by including another chiral Majorana loop, labeled by $\gamma^{\prime}$, which couples only to the original loop $\gamma$ with tunneling amplitude $t$ (see Fig. \ref{fig:l11}b). This extended setup can be used to model, for example, the effect of an additional point contact formed by confining the path of the original $\chi$MM, or scattering upon two coupled MBSs. The computation for the extended setup requires a minimal effort, which amounts to replacing $\varphi$ in Eq. \eqref{eq:r1} with $\varphi_{\mbox{\scriptsize ext}}$, defined as
\begin{align}\label{eq:phi_ext}
  \varphi_{\mbox{\scriptsize ext}} = \varphi + \arg(\frac{r-e^{i\varphi^{\prime}}}{1-r e^{i\varphi^{\prime}}}),
\end{align}
where ${\varphi}^{\prime} = EL^{\prime}/\hbar v_M + \pi + {\phi}^{\prime}$ with ${\phi}^{\prime}$ the magnetic flux inside $\gamma^{\prime}$, and $r = \sqrt{1-t^2}$ is the local reflection amplitude between the two chiral Majorana loops. The expressions for the average current and the noise power are unchanged from Eqs. \eqref{eq:Il11} and \eqref{eq:Pl11}.

\begin{figure}
  \centering
  \includegraphics[width=0.4\textwidth]{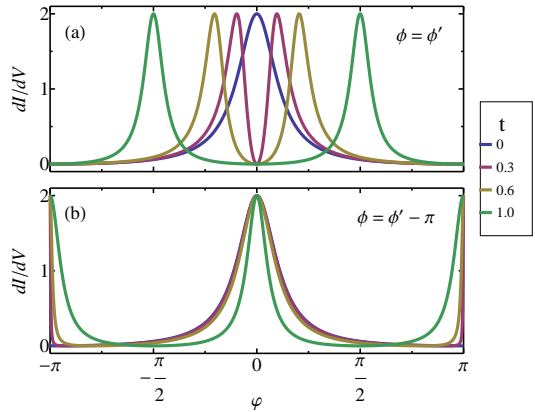}\\
  \caption{Differential conductance, $dI/dV$ in units of $e^2/h$,  as a function of $\varphi = 2\pi eV/E_l + \pi + \phi$ with $\phi \equiv n_v\pi$ for the setup in Fig. \ref{fig:l11}b. We set $\phi = \phi^\prime$ in the upper panel and set $\phi = \phi^\prime-\pi$ in the lower panel; for both panels $L=L^\prime$ is assumed and $r_0$ is fixed to be $0.7$. In the upper panel the resonance peak splits into two with their distance proportional to the energy splitting of the two coupled chiral Majorana loops; in the lower panel the resonance peak persists whenever $\varphi$ is an integer multiple of $\pi$ as long as the coupling between the two chiral Majorana loops is finite.}\label{fig:Gl11_2}
\end{figure}

For simplicity we consider $\phi = \phi^{\prime} = \pi$ and $L=L^{\prime}$. When the coupling between the two loops are turned off, we know from previous discussions that a resonance-led total Andreev reflection occurs at $E=0$ (more generally, at $E=nE_l$ with $n$ an integer; we will focus on the energy range around zero). Upon turning on the coupling -- even with infinitesimal $t$ -- the resonance at zero energy is immediately switched off because $\varphi_{\mbox{\scriptsize ext}}$ has a $\pi$ shift from $\varphi$ and hence $r_1$ changes sign. Indeed, since the energy spectrum of the coupled loops is given by $e^{i\varphi_{\mbox{\scriptsize ext}}} = 1$ (which is also equivalent to the resonance condition $r_1 = -1$), we find that the resonance shifts to $E \simeq \pm(E_l/2\pi)\,t$ when $t\ll1$. This is in close analogy to a system in which two MBSs (e.g. at the two ends of a chiral p-wave superconducting wire) are coupled and lifted from zero energy \cite{pikulin_topological_2011}. It deserves to be mentioned that if the current case is modified by setting one of the phases $\phi$ or $\phi^{\prime}$ to be $0$ and the other $\pi$, then the resonance occurs at zero energy regardless of the coupling strength between the loops as long as it is finite. Differential conductances for both situations, $\phi^{\prime} - \phi = 0$ or $\pi$, are plotted in Fig. \ref{fig:Gl11_2}.

\subsection{Fabry-Perot interferometer: $N=1$}
\label{ssec:FPI1}

\begin{figure}
  \centering
  \includegraphics[width=0.4\textwidth]{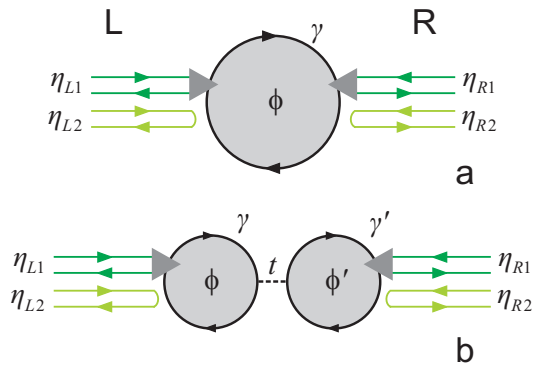}\\
  \caption{Majorana scattering channels for Fabry-Perot interferometers considered in Sec. \ref{ssec:FPI1}: the upper panel case (a) can be extended to the lower panel case (b) to model the presence of an additional point contact or two coupled MBSs (see text for details).}\label{fig:l21}
\end{figure}

Our second example deals with the Fabry-Perot interferometer composed of two single-mode normal leads (labeled by $L$ and $R$) probing a chiral Majorana loop simultaneously. We assume that the two normal leads are mediated only by the $\chi$MM and other tunneling processes are negligible. Fig. \ref{fig:l21}a illustrates the effective scattering in the Majorana language, and the corresponding Majorana scattering matrix is given by
\begin{equation}
\label{eq:Sl21}
  S_M =
  \begin{pmatrix}
    r_{L1} & 0 & t_{LR} & 0 \\
    0 & r_{L2} & 0 & 0 \\
    t_{RL} & 0 & r_{R1} & 0 \\
    0 & 0 & 0 & r_{R2}
  \end{pmatrix},
\end{equation}
where $r_{L2} = r_{R2} = 1$, and
\begin{align}
  &r_{L1} = Z(r_{L0}-r_{R0}e^{i\varphi}), \label{eq:rL1}\\
  &r_{R1} = Z(r_{R0}-r_{L0}e^{i\varphi}), \label{eq:rR1}\\
  &t_{LR} =-Zt_{L0}t_{R0}e^{i\varphi}, \label{eq:tLR}\\
  &t_{RL} =-Zt_{L0}t_{R0}, \label{eq:tRL}\\
  &Z\equiv (1-r_{L0}r_{R0}e^{i\varphi})^{-1}, \label{eq:Z}
\end{align}
with the subscript ${0}$ meaning that the reflection/transmission amplitude is local at the junction (corresponding to Eq. \eqref{eq:sm_1}, $r_0 = \cos\theta$, $t_0 = \sin\theta$, for two leads separately). Note that we have chosen a gauge such that $\varphi$ belongs entirely to the lower arm.

\begin{figure}
  \centering
  \includegraphics[width=0.4\textwidth]{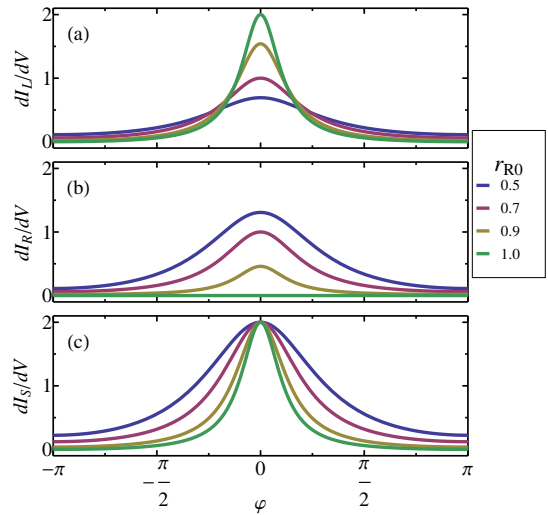}
  \caption{Differential conductance measured at different contacts ($L,R,S$) and in units of $e^2/h$,  as a function of $\varphi = 2\pi eV/E_l + (n_v+1)\pi$ for the Fabry-Perot interferometer in Fig. \ref{fig:l21}a, where $r_{L0}$ is fixed to be $0.7$ and the two normal contacts are equally biased. In general, resonance peaks (anti-resonance valleys) of the differential conductances at $\varphi = 0$ ($\varphi = \pm\pi$) persist as a robust feature. Asymmetric coupling at the two junctions ($r_{L0} \ne r_{R0}$) results in unbalanced resonance peaks in $dI_L/dV$ and $dI_R/dV$, while the resonance peak in $dI_S/dV$ remains quantized at $2e^2/h$. Particularly when one of the junctions is shut off (e.g. $r_{R0}=1$), the result returns to what is shown in Fig. \ref{fig:Gl11}, where the resonant-Andreev-reflection is completely local. This plot is again essentially a reproduction of the results in Ref. \cite{law_majorana_2009}, presented here for the purpose of comparison with later results.}\label{fig:Gl21}
\end{figure}

The average current and the zero-frequency noise power are given by
\begin{align}
  &I_L = \frac{e}{h}\int_{E\geq 0} dE\, \bigl[1-\Re(r_{L1}r_{L2}^*)\bigr]\delta{f}_{L}, \label{eq:Il21}\\
  &P_{LL}= \frac{e^2}{h}\int_{E\geq 0} dE\, \Bigl\{2\bigl[1-\Re(r_{L1}r_{L2}^*)\bigr] \Theta_{L} \nonumber\\
    &\qquad\qquad\qquad\qquad + \Im(r_{L1}r_{L2}^*)^2\delta{f}_L^2 \nonumber\\
    &\quad + \frac{1}{2}T_{LR}\bigl[\Theta_{R} - \Theta_{L} + \frac{1}{2}\sum\limits_{a,b=e,h}({f}_{La}-{f}_{Rb})^2\bigr]\Bigr\}, \label{eq:Pal21}\\
  &P_{LR}= P_{RL} \nonumber\\
    &= \frac{e^2}{h} \int_{E\ge0} dE\, (-\frac{1}{2})\Re(t_{LR}r_{L2}^*t_{RL}r_{R2}^*)\delta{f}_{L}\delta{f}_{R}, \label{eq:Pcl21}
\end{align}
where $T_{LR} = t_{LR}^*t_{LR}$. $I_R$ and $P_{RR}$ can be obtained by interchanging subscripts $L$ and $R$ in the expressions for $I_L$ and $P_{LL}$.

In the presence of multiple normal contacts, it is particularly interesting to look at the average current $I_S$ flowing through the grounded superconducting contact. Assuming the two normal contacts here are equally biased ($\delta\mu_L = \delta\mu_R = eV$), $I_S$ is simply given by
\begin{align}
  &I_S= I_L+I_R \nonumber\\
  &= \frac{e}{h}\int_{E\geq 0} dE\,\bigl[2 - (r_{L0}+r_{R0}) \Re(\frac{1-e^{i\varphi}}{1-r_{L0}r_{R0}e^{i\varphi}}) \bigr] \delta{f}, \label{eq:ISl21}
\end{align}
where we have used the explicit expressions of the reflection amplitudes. An immediate observation based on  the above result is that the resonance peak in the differential conductance $dI_S/dV$ at $\varphi=0$ remains quantized at $2e^2/h$ regardless of the coupling of the two normal leads to the $\chi$MM (see Fig. \ref{fig:Gl21}).

\subsubsection{Cross-correlation as two-MF-exchange effect}

Formally the expression \eqref{eq:Il21} for $I_L$ (or $I_R$) is the same as Eq. \eqref{eq:Il11} in the single lead case, meaning that one contact does not contribute to the average current at the other contact regardless of the electron/hole occupation in either contact. Naively this can be seen as a result of the charge neutrality of the $\chi$MM that bridges two contacts. A more precise interpretation, however, is that the Majorana scattering states sourced from different contacts (e.g. $\eta_{L1}$ transmitted from contact $R$ and $\eta_{L2}$) are not phase coherent when only single-particle scattering is considered, therefore cannot lead to finite average current according to our discussion in Sec. \ref{ssec:l11}. This statement will be clear after we further examine the cross-correlation between the two contacts.

The charge-current cross-correlator $P_{LR}$  (or $P_{RL}$), given by Eq. \eqref{eq:Pcl21}, is manifestly non-vanishing at low temperature and low bias if, and only if, both contacts are biased ($\delta{f}_{L,R}(E) \ne 0$) and exchange of Majorana states is allowed ($t_{LR}t_{RL} \ne 0$). This is despite the fact that the mediating $\chi$MM carries no charge current by itself. Indeed, the non-vanishing cross-correlation here is entirely due to those scattering events that involve two-particle exchange and permit coherence between scattering states from different sources, and hence is entirely the ``exchange noise". Such an exchange contribution is irrelevant for the average current but present in noise. Meanwhile, the absence of other types of noise (e.g. thermal noise, partition noise) in the cross-correlator again highlights the essential role that the interference between Majorana modes plays in charge transport.

In contrast to the cross-correlation, the auto-correlation for contact $L$, given by Eq. \eqref{eq:Pal21}, contains both the equilibrium thermal contribution (the first term in the kernel) and nonequilibrium contributions (the rest two terms). Contact $R$ manifests itself in $P_{LL}$ only away from equilibrium (except for modifying $r_{L0}$ to $r_{L1}$), which is clearly different from a conventional normal-metal-superconductor-normal-metal device\cite{anantram_current_1996}.

\subsubsection{The zero-temperature low-bias limit}

\begin{figure}
  \centering
  \includegraphics[width=0.45\textwidth]{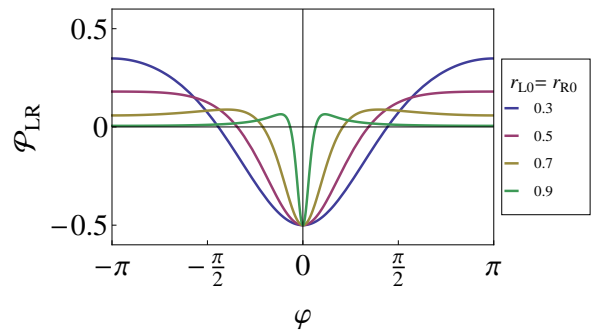}\\
  \caption{The kernel $\mathcal{P}_{LR}(E)$ of the integral for the cross-correlator $P_{LR}$ in Eq. \eqref{eq:Pcl21}, excluding the factor due to Fermi distribution functions, as a function of $\varphi(E)$. The coupling between the normal leads and the $\chi$MM is assumed to be symmetric. Situations with various coupling strength are plotted. Between resonance ($\varphi = 0$) and  anti-resonance ($\varphi = \pm\pi$) points (cf. Fig. \ref{fig:Gl21}), $\mathcal{P}_{LR}$ changes sign.}\label{fig:Pl21}
\end{figure}

At zero-temperature and low bias ($eV \ll E_l$, equal for both normal contacts), the average current and the zero-frequency noise power can be reduced to \cite{law_majorana_2009}:
\begin{align}
  &\hspace{-3mm}\mbox{for even $n_v$:} \nonumber\\
  &\hspace{-1mm} I_L = I_R = \frac{1}{2}I_S = \frac{e^2 V}{h}(1-r_1)\,, \label{eq:Il21_e}\\
  &\hspace{-1mm} P_{LL} = P_{RR} = P_{LR} = P_{RL} =\frac{e^3V}{2h}(1-r_1^2)\,, \label{eq:Pl21_e}\\
  &\hspace{-1mm} r_1 \equiv r_{L1} = r_{R1} = \frac{r_{L0}+r_{R0}}{1+r_{L0}r_{R0}}\,; \label{eq:r1l21_e}\\
  &\hspace{-3mm}\mbox{for odd $n_v$:} \nonumber\\
  &\hspace{-1mm} I_L = \frac{e^2 V}{h}(1-r_1),\, I_R = \frac{e^2 V}{h}(1+r_1),\, I_S = \frac{2e^2}{h} V, \label{eq:l21_o}\\
  &\hspace{-1mm} P_{LL} = P_{RR} = -P_{LR} = -P_{RL} =\frac{e^3V}{2h}(1-r_1^2)\,, \label{eq:Pl21_o}\\
  &\hspace{-1mm} r_1 \equiv r_{L1} = -r_{R1} = \frac{r_{L0}-r_{R0}}{1-r_{L0}r_{R0}}\,. \label{eq:r1l21_o}
\end{align}
The Fano factors, for individual normal leads defined as $F_\alpha = P_{\alpha\alpha}/(eI_\alpha)$, and for the total noise power (or equivalently, for the superconducting contact) defined as $F = (\sum_{\alpha,\beta}P_{\alpha\beta})/(e\sum_{\alpha}I_\alpha)$, where $\alpha,\beta = L,R$, are given by
\begin{align}
  &\hspace{-3mm}\mbox{for even $n_v$:}\quad F_L = F_R = \frac{1+r_1}{2},\; F = 1+r_1\,; \label{eq:FFl21_e}\\
  &\hspace{-3mm}\mbox{for odd $n_v$:}\quad F_L = \frac{1+r_1}{2},\; F_R = \frac{1-r_1}{2},\; F = 0\;. \label{eq:FFl21_o}
\end{align}

We see that the even/odd $n_v$ cases represent rather distinct processes. For even $n_v$ the conductance and Fano factors of both contacts $L$ and $R$ are equal. This indicates that the statistics of the electron transport is in fact identical at these leads and the process consists of a random sequence of crossed Andreev reflection processes \cite{law_majorana_2009}. In contrast for $n_v$ odd, electrons are drawn at different rates from both contacts but such as to generate a perfect stream of Cooper pairs entering the superconductor noiselessly ($F=0$). Meanwhile, the cross-correlators change from being positive for even $n_v$ to being negative for odd $n_v$, indicating increasing importance of local Andreev reflections.

For a more general picture of the scattering processes beyond the low-bias limit, we plot in Fig. \ref{fig:Pl21} the cross-correlation noise power spectrum (the kernel of the integral in Eq. \eqref{eq:Pcl21} excluding the factor due to Fermi distribution functions) $\mathcal{P}_{LR} \equiv (-\frac{1}{2})\Re(t_{LR}r_{L2}^*t_{RL}r_{R2}^*)$ as a function of $\varphi$. By using the explicit expressions of the scattering amplitudes, one finds clearly that $\mathcal{P}_{LR}$ changes sign between resonance ($\varphi = 0 \mod 2\pi$) and  anti-resonance ($\varphi = \pi \mod 2\pi$) points (cf. Fig. \ref{fig:Gl21}). The interpretation for this sign change is in fact the same as our preceding discussion for the low-bias cases.

\subsubsection{Extended setups}

\begin{figure}
  \centering
  \includegraphics[width=0.45\textwidth]{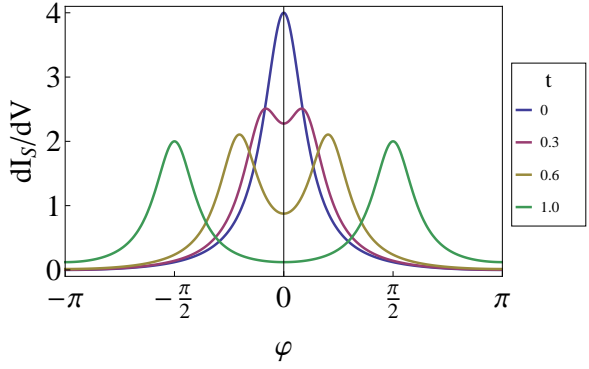}
  \caption{Differential conductance measured at the superconducting contact, $dI_S/dV$ in units of $e^2/h$, as a function of $\varphi = 2\pi eV/E_l + \pi + \phi$ with $\phi \equiv n_v\pi$ for the extended Fabry-Perot interferometer in Fig. \ref{fig:l21}b. We assume $\phi = \phi^\prime$, $L=L^\prime$ and $r_{L0}=r_{R0}=0.7$. Situations with various $t$ between the two loops are plotted. When $t=0$, this setup is simply two copies of the single-lead setup in Fig. \ref{fig:l11}; when $t=1$, this setup reduces to the original Fabry-Perot interferometer in Fig. \ref{fig:l21}a, with $\varphi+\varphi^\prime-\pi$ playing the role of $\varphi$ in the original setup. For the case of $\phi = \phi^\prime -\pi$ and $L=L^\prime$, $I_S$ does not depend on $t$ at symmetric coupling ($r_{L0}=r_{R0}$), and its expression reduces to Eq. \eqref{eq:ISl21} (see also Fig. \ref{fig:Gl21}c) with $\varphi$ substituted by $2\varphi$. The plot shows that quite generally, $dI_S/dV$ peaks at resonance (upon eigenstates of the coupled loops) points, and is suppressed at anti-resonance points.}\label{fig:Gl21_2}
\end{figure}

\begin{figure}
  \centering
  \includegraphics[width=0.45\textwidth]{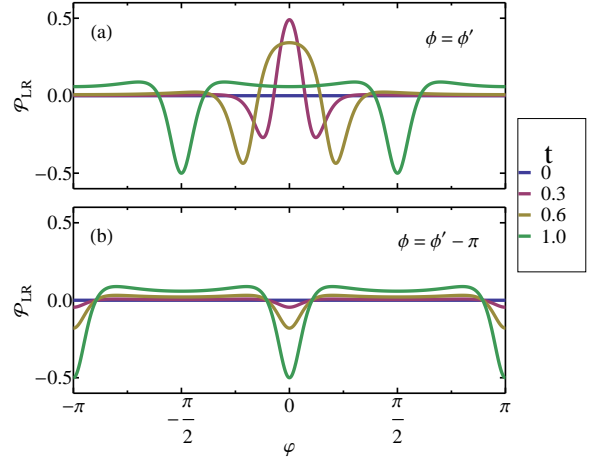}
  \caption{Cross-correlation noise power spectrum $\mathcal{P}_{LR}$ as a function of $\varphi(E)$ for the extended Fabry-Perot interferometer in Fig. \ref{fig:l21}b. In the upper panel we assume $\phi = \phi^\prime$, and in the lower panel $\phi = \phi^\prime-\pi$; for both panels we take $L=L^\prime$ and $r_{L0}=r_{R0}=0.7$. Situations with various transmission amplitudes $t$ between the two loops are plotted. Clearly between resonance and  anti-resonance points (cf. Fig. \ref{fig:Gl21_2}), $\mathcal{P}_{LR}$ changes sign.}\label{fig:Pl21_2}
\end{figure}

As in the single lead case discussed in Sec. \ref{ssec:l11}, the present setup can be extended in the same way by including another chiral Majorana loop which couples to the original one (see Fig. \ref{fig:l21}b). For the extended setup, the following substitutions for the scattering amplitudes need to be made
\begin{align}
  &r_{L1} = Z\bigl[r_{L0}+r_{R0}e^{i(\varphi+\varphi^\prime)}-r(e^{i\varphi}+r_{L0}r_{R0}e^{i\varphi^\prime})\bigr], \label{eq:rL1_2}\\
  &r_{R1} = Z\bigl[r_{R0}+r_{L0}e^{i(\varphi+\varphi^\prime)}-r(e^{i\varphi^\prime}+r_{L0}r_{R0}e^{i\varphi})\bigr], \label{eq:rR1_2}\\
  &t_{LR} = Zt_{L0}t_{R0}t\,e^{i(\varphi+\varphi^\prime)}, \label{eq:tLR_2}\\
  &t_{RL} = -Zt_{L0}t_{R0}t, \label{eq:tRL_2}\\
  &Z\equiv \bigl[t^2+(r-r_{L0}e^{i\varphi})(r-r_{R0}e^{i\varphi^\prime})\bigr]^{-1}. \label{eq:Z_2}
\end{align}
Note that the above amplitudes reduce to Eq. \eqref{eq:r1} at $r=\sqrt{1-t^2}=1$, and to Eqs. \eqref{eq:rL1}-\eqref{eq:tRL} at $r=0$, $t=1$. We plot the differential conductance $dI_S/dV$ and the cross-correlation noise power spectrum $\mathcal{P}_{LR}$ for the present case in Fig. \ref{fig:Gl21_2} and \ref{fig:Pl21_2}, respectively. Similar to our previous discussions, the coupling between the two loops (assumed again to be of the same circumferences) leads to split resonance peaks if $\phi = \phi^\prime$, or a halved period if $\phi = \phi^\prime - \pi$, for the differential conductance $dI_S/dV$; the sign of $\mathcal{P}_{LR}$ shows consistently changes between resonance and anti-resonance points upon eigenstates of the coupled loops.

\subsection{A single normal lead probing a chiral Majorana loop: $N>1$}
\label{ssec:l1n}

\begin{figure}
  \centering
  \includegraphics[width=0.3\textwidth]{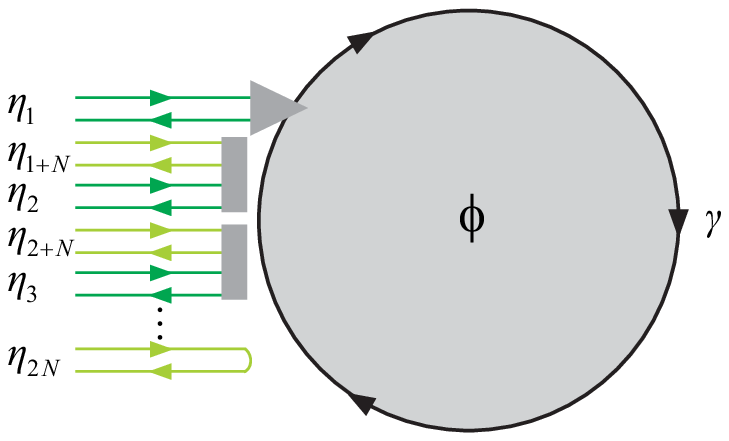}\\
  \caption{Effective scattering for a setup with a multi-mode ($N>1$) normal lead probing a chiral Majorana loop. Only the (single) Majorana mode coupled to the chiral Majorana loop can ``feel" the phase picked up in the loop.}\label{fig:l1n}
\end{figure}

Our previous two examples deal with single-mode ($N=1$) leads, which are simple but nevertheless reveal most of the interesting physics. The real strength of the decompositions discussed in Sec. \ref{sec:contact}, however, becomes obvious only in dealing with multi-mode ($N>1$) leads. In this section we start again with a single-lead setup similar to that in Sec. \ref{ssec:l11} but with $N>1$ (see Fig. \ref{fig:l1n}).

According to Sec. \ref{ssec:mcdec}, the Majorana scattering matrix for this setup can be written as
\begin{equation}\label{eq:Sl1n}
  S_M =
  \begin{pmatrix}
    r_1  &   & & & \\
         & \begin{matrix}
            r_2 & -t_2\\
            t_2 & r_2
            \end{matrix} & & \mbox{\LARGE{0}} & \\
         & &  \ddots & & \\
         & \mbox{\LARGE{0}} & &\begin{matrix}
               r_{N} & -t_{N}\\
               t_{N} & r_{N}
              \end{matrix} & \\
         & & & & r_{N+1}
  \end{pmatrix},
\end{equation}
where $r_{N+1} = 1$, $r_1$ is given by Eq. \eqref{eq:r1}, and $r_i = \cos{\theta_i}$ and $t_i = \sin{\theta_i}$ with $1<i\le N$ in accordance with Eq. \eqref{eq:sm_n}. Note that the basis for this scattering matrix is reordered from the original one defined by Eq. \eqref{eq:eta}, to be $\bm{\eta} = (\eta_1,\eta_{1+N},\eta_2,\eta_{2+N},...,\eta_N,\eta_{N+N})^T$, where one should keep in mind that $\eta_i$ and $\eta_{i+N}$ are the partners associated with the $i$-th original mode as given by Eqs. \eqref{eq:eta_i} and \eqref{eq:eta_in}.

In terms of the elements of this scattering matrix the average current and the zero-frequency noise power (auto-correlator) are:
\begin{align}
  I &= \frac{e}{h}\int_{E\geq 0} dE\, \sum\limits_{i=1}^{N} \bigl[1-\Re(r_ir_{i+1}^*)\bigr]\delta{f}, \label{eq:Il1n}\\
  P &= \frac{e^2}{h}\int_{E\geq 0} dE\, \Bigl\{\sum\limits_{i=1}^{N}2\bigl[1-\Re(r_ir_{i+1}^*)\bigr]\Theta \nonumber\\
    &\qquad + \sum\limits_{i=1}^{N}\frac{1}{2}\bigl[1-\Re(r_ir_{i+1}^*)^2+\Im(r_ir_{i+1}^*)^2\bigr]\delta{f}^2 \nonumber\\
    &\qquad + \sum\limits_{i=2}^{N}\Re(t_ir_{i-1}^*t_ir_{i+1}^*)\delta{f}^2\Bigr\}. \label{eq:Pl1n}
\end{align}

Clearly the total average current is a sum of contributions from all $N$ modes individually: $r_i$ and $r_{i+1}$ are reflection amplitudes for $\eta_i$ and $\eta_{i+N}$ and their interference term $\Re(r_ir_{i+1}^*)$ registers the outgoing current contributed by mode $i$; scattering between different modes (i.e. $t_i$) does not appear directly in the average current similar to the case of scattering between contacts in a Fabry-Perot interferometer. The noise power, by contrast, contains not only the auto-correlation of each individual mode (the first two terms in Eq. \eqref{eq:Pl1n}; cf. Eq. \eqref{eq:Pl11}), but also the cross-correlation between different modes (the last term in Eq. \eqref{eq:Pl1n}; cf. Eq. \eqref{eq:Pcl21}).

In general, $\{r_i:\;1<i\le N\}$ are parameters determined specifically by the details of the junction between the normal lead and the TSC, and do not necessarily lead to quantized conductance regardless of the value of $\varphi$. This is certainly an important consequence of the presence of multiple modes in the normal lead. An interesting exception \cite{wimmer_quantum_2011}, however, is the zero-temperature low-bias conductance when $N=2$ and $n_v$ is odd, where one finds $r_1 = -r_3 = -1$, thus $G = 2e^2/h$ and $P=0$. Wimmer et al. \cite{wimmer_quantum_2011} have proposed this exceptional case to be a robust transport signature for detecting a MBS, with which a zero-energy eigenstate of the chiral Majorana loop can be identified \footnote{Indeed, such robustness is of a more fundamental origin, namely the B\'{e}ri degeneracy \cite{beri_dephasing-enabled_2009}, which we will discuss in details in our following work, using the Majorana picture similar to this paper.}.

\subsection{$K$ multi-mode normal leads coupled to a chiral Majorana loop}
\label{ssec:lkn}

\begin{figure}
  \centering
  \includegraphics[width=0.4\textwidth]{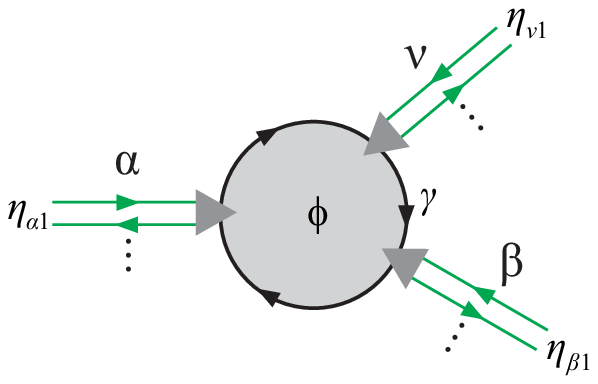}\\
  \caption{Effective scattering channels for a generic Fabry-Perot-type setup with $K$ multi-mode normal leads coupled to a single chiral Majorana loop. In each lead, only one Majorana mode is coupled to the chiral Majorana loop, and hence is involved in transmission between contacts (any direct tunneling between contacts is ignored); the other Majorana modes (omitted in this illustration) are locally, pairwise reflected.}\label{fig:lkn}
\end{figure}

As the final example for the Fabry-Perot-type interferometers making use of a single $\chi$MM, we discuss a generic setup with $K$ normal contacts each admitting $N_\alpha$ ($\alpha=1,...,K$) transmission modes in the lead coupled to the $\chi$MM (see Fig.\ref{fig:lkn}). The Majorana scattering matrix for this setup is given by
\begin{align}
  &S_M =
  \begin{pmatrix}
    s_{11} & s_{12} & \cdots & s_{1K} \\
    s_{21} & s_{22} & \cdots & s_{2K} \\
    \vdots & \vdots & \ddots & \vdots \\
    s_{K1} & s_{K2} & \cdots & s_{KK} \\
  \end{pmatrix} \quad\mbox{with} \label{eq:Slkn} \\
  &s_{\alpha\alpha}= \nonumber\\
  &\begin{pmatrix}
    r_{1(\alpha)}  &   & & & \\
         & \begin{matrix}
            r_{2(\alpha)} & -t_{2(\alpha)}\\
            t_{2(\alpha)} & r_{2(\alpha)}
            \end{matrix} & & \mbox{\LARGE{0}} & \\
         & &  \ddots & & \\
         & \mbox{\LARGE{0}} & &\begin{matrix}
               r_{N_\alpha(\alpha)} & -t_{N_\alpha(\alpha)}\\
               t_{N_\alpha(\alpha)} & r_{N_\alpha(\alpha)}
              \end{matrix} & \\
         & & & & r_{N_\alpha+1(\alpha)}
  \end{pmatrix}, \label{eq:Saa} \\
  &s_{\alpha\beta}=
  \begin{pmatrix}
    t_{\alpha\beta}  &  \\
         & \mbox{\LARGE{0}}
  \end{pmatrix}, \quad(\alpha\ne\beta) \label{eq:Sab}
\end{align}
where
\begin{align}
  &r_{1(\alpha)}= Z(r_{0(\alpha)}-e^{i\varphi}\prod_{\beta\ne\alpha}r_{0(\beta)}), \label{eq:r1a} \\
  &t_{\alpha\beta}= -Z t_{0(\alpha)}t_{0(\beta)}e^{i\varphi_{\alpha\beta}},\;
  \left(\varphi_{\alpha\beta} = \Bigl\{
  \begin{array}{ll}
    \varphi, & \alpha<\beta \\
    0, & \alpha>\beta
  \end{array}\right) \label{eq:tab} \\
  &Z\equiv (1-e^{i\varphi}\prod_{\beta}r_{0(\beta)})^{-1}, \label{eq:ZZ}
\end{align}
and $r_{i(\alpha)}$ and $t_{i(\alpha)}$ with $i>1$ take the values according to Eq. \eqref{eq:sm_n} for contact $\alpha$. The subscript ${0}$ again means that the reflection/transmission amplitude is local at the junction and contains no $\varphi$-dependence. The basis here for each contact is ordered the same way as in \eqref{eq:Sl1n}.

The average current and the zero-frequency noise powers are given by
\begin{align}
  &I_\alpha= \frac{e}{h}\int_{E\geq 0} dE\, \sum\limits_{i=1}^{N_\alpha} \bigl[1-\Re(r_{i(\alpha)}r_{i+1(\alpha)}^*)\bigr]\delta{f}_{\alpha}, \label{eq:Ilkn}\\
  &P_{\alpha\alpha}= P_{\alpha\alpha}^{(0)} + \frac{e^2}{h}\int_{E\geq 0} dE\,\cdot \nonumber\\ &\quad\sum\limits_{\beta\ne\alpha}\frac{1}{2}T_{\alpha\beta}\bigl[\Theta_{\beta} - \Theta_{\alpha} - \delta{f}_{\alpha}^2 + \frac{1}{2}\sum\limits_{a,b=e,h}({f}_{{\alpha}a}-{f}_{{\beta}b})^2\bigr], \label{eq:Palkn}\\
  &P_{\alpha\beta} = \frac{e^2}{h} \int_{E\ge0} dE\, (-\frac{1}{2})\Re(t_{\alpha\beta}r^*_{2(\alpha)}t_{\beta\alpha}r^*_{2(\beta)})\delta{f}_{\alpha}\delta{f}_{\beta}, \nonumber\\
  &\hspace{60mm}(\alpha\ne\beta) \label{eq:Pclkn}
\end{align}
where $P_{\alpha\alpha}^{(0)}$ takes the form of the single-lead auto-correlator \eqref{eq:Pl1n}.

A remarkable fact in these results is that in the presence of many modes, the $\varphi$-dependence of the average current or the auto-correlators, carried only by $r_{1(\alpha)}$ and $t_{\alpha\beta}$, can be overwhelmed by the large $\varphi$-independent contributions from $r_{i(\alpha)}$ and $t_{i(\alpha)}$ ($i>1$) at imperfect interfaces between normal leads and the TSC; the cross-correlators, however, preserve the $\varphi$-dependent terms as their sole contributions, hence will optimally manifest the information embedded in the transmission of $\chi$MM.

Without performing further quantitative analysis for this example, we simply point out two recurrent messages in the above general results: the average current at one contact is \textit{only} contributed by the interference between pairs of Majorana modes in the corresponding lead, and the current cross-correlation between two contacts is \textit{purely} due to coherent exchange of two MFs sourced from the two contacts. These two messages together recapitulate the crucial role that interference between Majorana modes plays in charge transport.

\subsection{Mach-Zehnder and Hanbury Brown-Twiss interferometers}
\label{ssec:mz_hbt}

\begin{figure}
  \centering
  \includegraphics[width=0.48\textwidth]{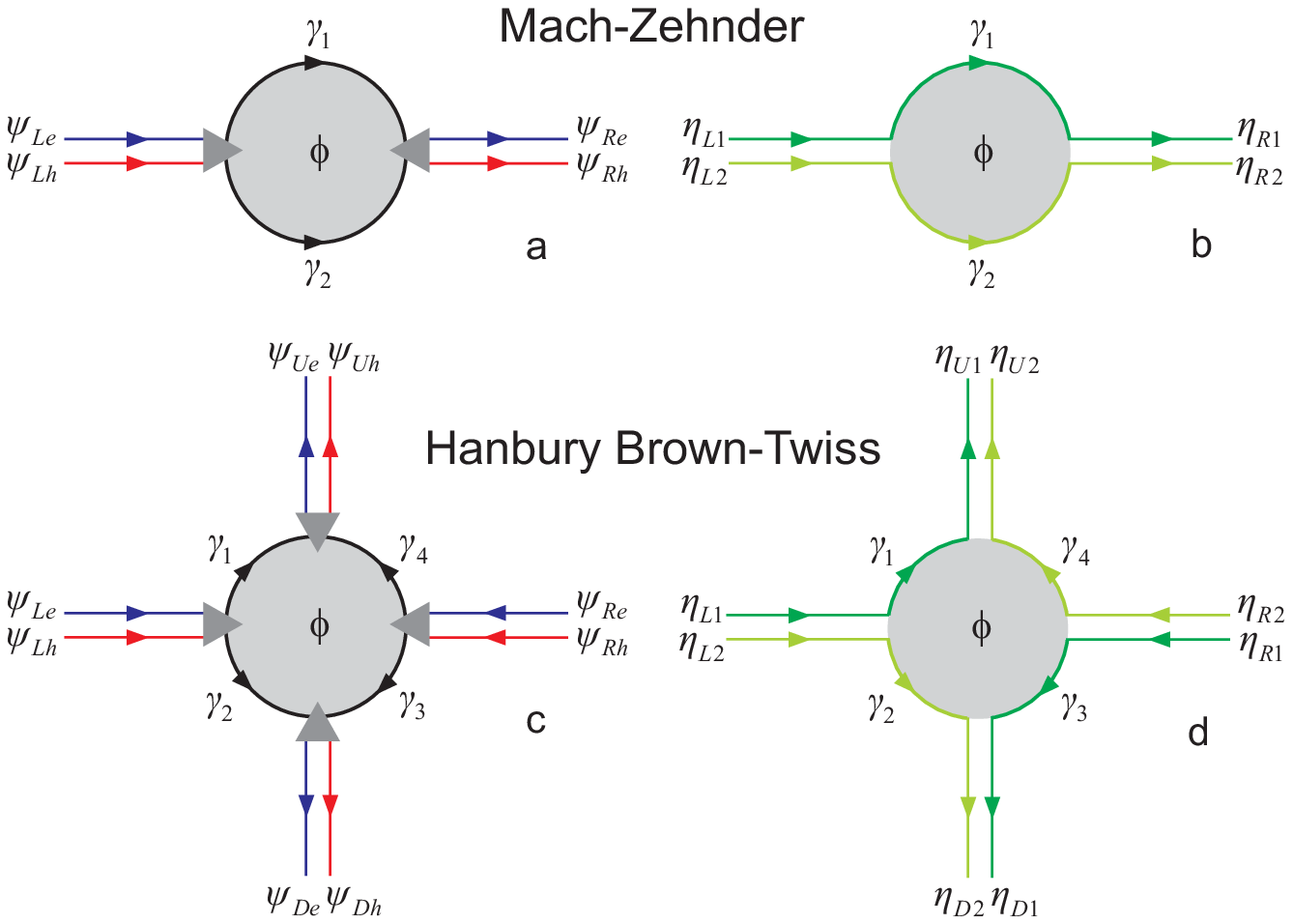}\\
  \caption{Illustrations of the scattering channels in a Mach-Zehnder interferometer (a,b) and a Hanbury Brown-Twiss interferometer (c,d), in the electron-hole basis for the normal part (a,c), and in the properly-chosen Majorana basis (b,d). In the latter basis, the Majorana scattering matrix for a junction between a normal channel and two $\chi$MMs is simply an identity matrix. In both setups the outgoing leads are connected to electron reservoirs that are grounded.}\label{fig:mz_hbt}
\end{figure}

So far we have been focusing on the junctions between normal leads and a single $\chi$MM, and the Fabry-Perot-type interferometers which comprise only such junctions. Alternatively another type of junctions, composed of two $\chi$MMs with opposite chirality and connected to a chiral normal channel, can be used to build interferometers of the Mach-Zehnder type \cite{fu_probing_2009, akhmerov_electrically_2009} or the Hanbury Brown-Twiss type \cite{struebi_interferometric_2011} (see Fig. \ref{fig:mz_hbt}).

The junctions between two $\chi$MMs and a chiral normal channel are reflectionless and can be described by two-times-two scattering matrices \cite{fu_probing_2009, akhmerov_electrically_2009}, where the two components at the normal side correspond to electron and hole modes or two Majorana modes depending on the basis. Assuming again the scattering matrix at the junction is energy-independent at low enough energy, the Majorana scattering matrix belongs to $SO(2)$ owing to the PHS. It follows that by employing the same strategy as presented in Sec. \ref{ssec:gen}, the canonical form of the scattering matrix, in the properly chosen Majorana basis, is simply an identity matrix. In other words, the junction effectively splits or joins, without mixing, two $\chi$MMs. This certainly simplifies significantly our discussions which we present below in terms of specific examples.

\subsubsection{Mach-Zehnder interferometer}

The Mach-Zehnder interferometer illustrated in Fig. \ref{fig:mz_hbt}a has been investigated by Fu and Kane \cite{fu_probing_2009}, and Akhmerov, Nilsson and Beenakker \cite{akhmerov_electrically_2009}. Here we simply reformulate the problem in the Majorana language we have been speaking so far, to echo some general features of Majorana transport that have been discussed in the previous part of this paper.

The Majorana scattering matrix for the Mach-Zehnder interferometer(see Fig. \ref{fig:mz_hbt}b), defined as
\begin{align}\label{eq:smdef_mz}
  \begin{pmatrix}
    \eta_{R1} \\
    \eta_{R2} \\
  \end{pmatrix}
   = S_M
   \begin{pmatrix}
    \eta_{L1} \\
    \eta_{L2} \\
  \end{pmatrix},
\end{align}
is effectively given by
\begin{align}\label{eq:sm_mz}
  S_M =
  \begin{pmatrix}
    t_1 & 0 \\
    0 & t_2 \\
  \end{pmatrix},
\end{align}
where we can choose a specific gauge such that $t_1 = e^{i\varphi}$ with $\varphi = (n_v+1) \pi + 2\pi E/E_l$, and $t_2 = 1$.

The average current and zero-frequency noise power can be easily obtained:
\begin{align}
  &I_L = \frac{e}{h}\int_{E\geq 0} dE \; \delta{f}_{L}, \label{eq:IL_mz}\\
  &I_R = -\frac{e}{h}\int_{E\geq 0} dE \; \Re(t_1t_2^*)\; \delta{f}_{L}, \label{eq:IR_mz}\\
  &I_S = \frac{e}{h}\int_{E\geq 0} dE \; \bigl[1-\Re(t_1t_2^*)\bigr]\;\delta{f}_{L}, \label{eq:IS_mz}\\
  &P_{LL} = \frac{e^2}{h}\int_{E\geq 0} dE \; \Theta_L, \label{eq:PLL_mz} \\
  &P_{RR} = \frac{e^2}{h}\int_{E\geq 0} dE \; \bigl[\Theta_L + \Im(t_1t_2^*)^2\,\delta{f}_L^2\bigr], \label{eq:PRR_mz} \\
  &P_{LR} = P_{RL} = -\frac{e^2}{h}\int_{E\geq 0} dE \; \Re(t_1t_2^*)\;\Theta_L\,, \label{eq:Pc_mz}
\end{align}
where $\Re(t_1t_2^*) = \cos\varphi$ and $\Im(t_1t_2^*) = \sin\varphi$ when the explicit expressions of $t_1$ and $t_2$ are taken.

Clearly the outgoing current in Eq. \eqref{eq:IR_mz} appears to be the interference between two $\chi$MMs -- the transmission ``probability" $\Re(t_1t_2^*)$ has a physically identical interpretation as the reflection ``probability" $\Re(r_1r_2^*)$ appearing in Eq. \eqref{eq:Il11}. An obvious difference, however, between the present setup and a Fabry-Perot-type interferometer is that contact $L$ here indeed provides coherent sources for the MF pairs --instead of single ones-- that are drained by contact $R$. As a direct consequence, the cross-correlator in Eq. \eqref{eq:Pc_mz} involves only the thermal fluctuation in contact $L$, while all the other contributions (e.g. partition or exchange noise) are absent due to the one-way nature of the present setup.

\subsubsection{Hanbury Brown-Twiss interferometer}

The Hanbury Brown-Twiss interferometer shown in Fig. \ref{fig:mz_hbt}c has recently been investigated by Str\"{u}bi, Belzig, Choi, and Bruder \cite{struebi_interferometric_2011}. Here we reformulate the problem, again in the Majorana language.

The Majorana scattering matrix for the Hanbury Brown-Twiss interferometer(see Fig. \ref{fig:mz_hbt}d), defined as
\begin{align}\label{eq:smdef_hbt}
  \begin{pmatrix}
    \eta_{U1} \\
    \eta_{U2} \\
    \eta_{D1} \\
    \eta_{D2} \\
  \end{pmatrix}
   = S_M
   \begin{pmatrix}
    \eta_{L1} \\
    \eta_{L2} \\
    \eta_{R1} \\
    \eta_{R2} \\
  \end{pmatrix},
\end{align}
is effectively given by
\begin{align}\label{eq:sm_hbt}
  S_M =
  \begin{pmatrix}
    t_{UL} & 0 & 0 & 0 \\
    0 & 0 & 0 & t_{UR} \\
    0 & 0 & t_{DR} & 0 \\
    0 & t_{DL} & 0 & 0 \\
  \end{pmatrix}.
\end{align}
In this case we can choose a specific gauge such that $t_{UL} = e^{i\varphi}$, and $t_{UR} = t_{DR} = t_{DL} = 1$.

The average current and zero-frequency noise power are given by
\begin{align}
  &I_{\alpha} = \frac{e}{h}\int_{E\geq 0} dE \;\delta{f}_{\alpha}\,,\quad (\alpha=L,R) \label{eq:Ia_hbt}\\
  &I_{\beta} = 0\,,\quad (\beta=U,D) \label{eq:Ib_hbt}\\
  &P_{\alpha\alpha} = \frac{e^2}{h}\int_{E\geq 0} dE \; \Theta_\alpha\,,\quad (\alpha=L,R) \label{eq:Paa_hbt} \\
  &P_{\beta\beta} = \frac{e^2}{h}\int_{E\geq 0} dE \;\bigl[\,\frac{1}{2}(\Theta_L + \Theta_R) \nonumber\\
  &\qquad\qquad + \frac{1}{4}\sum\limits_{a,b=e,h}({f}_{La}-{f}_{Rb})^2\,\bigr]\,,\quad (\beta=U,D) \label{eq:Pbb_hbt} \\
  &P_{UD} = P_{DU} \nonumber\\
  &= \frac{e^2}{h}\int_{E\geq 0} dE \; (-\frac{1}{2})\Re(t_{UL}t^*_{UR}t_{DR}t^*_{DL})\delta{f}_{L}\delta{f}_{R} \,, \label{eq:PUD_hbt}
\end{align}
and all other cross-correlators are zero. Taking the explicit expressions, we have $\Re(t_{UL}t^*_{UR}t_{DR}t^*_{DL}) = \cos\varphi$.

As pointed out by Str\"{u}bi \textit{et al.} \cite{struebi_interferometric_2011}, a remarkable fact in the present setup is that although the average current at the two drains $U$ and $D$ vanishes identically, the cross-correlation between the two contacts is in general non-zero. This fact, in the Majorana language, can be interpreted as follows: the MF pair in either drain are necessarily from different sources, and are therefore phase incoherent for single-particle scattering which the average current measures -- the incoherence between MF pairs implies vanishing average current; two-particle scattering allows coherence between MF pairs in both drains, and is included in the measurement of noise power -- the coherent injection of two electrons/holes from the two sources leads to non-vanishing cross-correlation between current in the two drains.

The scenario here bears close analogy to the $N=1$ Fabry-Perot interferometer discussed in Sec. \ref{ssec:FPI1}, where the average current contributed by a foreign contact is always zero but the cross-correlation between two contacts is in general non-vanishing. The underlying physics in both scenarios, as it has been for essentially all our examples, boils down to the central importance of interference in the charge transport with Majorana modes.

\section{Summary}

In this work we emphasized a scattering approach using a global Majorana basis. Both the normal state leads and the system which intrinsically carries chiral Majorana states are described in the same basis. As a consequence, from a purely technical particle scattering point of view, Majorana scattering problems are rather similar to scattering problems in normal mesoscopic structures. However, Majorana modes are neutral and carry neither charge nor (electric) current. In contrast to normal scattering problems the charge and current operators are not diagonal in the Majorana scattering states. Therefore, unlike in normal systems, current is not determined by transmission probabilities of Majorana fermions. We have shown that charge and current appear due to interference of pairs of Majorana modes. For averages of single particle quantities, like charge and current, to be non-vanishing, interfering Majorana modes must necessarily have originated in the same contact. In contrast, in two particle quantities, like the current noise, two-particle exchange permits interference of two Majorana particles even if they originate from different leads.
The physics of neutral excitations is certainly an important field of future research.  It can be expected that the theory of Majorana particles represents an instructive example of this development.

\section{Acknowledgement}

We thank C. W. J. Beenakker, P. W. Brouwer, A. R. Akhmerov, M. Wimmer, I. Martin and A. F. Morpurgo for discussions. M. B. thanks the KITP of Beijing and Santa Barbara for hospitality where part of this work was carried out. The work is supported by the Swiss National Science Foundation, the European Marie-Curie Training Network NanoCTM and the Swiss National Center of Competence in Research on Quantum Science and Technology.

\appendix
\section{Decomposition of a ($2N+1$)-dimensional orthogonal matrix}
\label{app:dec}
We prove in this appendix that for an arbitrary matrix $\mathcal{O}_N \in SO(2N+1)$, one can find a decomposition
\be
\label{eq:appdecO}
\mathcal{O}_N=
\begin{pmatrix}
1 & 0\\
0 & \mathcal{V}^T
\end{pmatrix}
\left(\begin{array}{ccccc}
    R(\theta_1) & &  & \mbox{\LARGE{0}}\\
    & \ddots & & \\
    & & R(\theta_{N}) & \\
    \mbox{\LARGE{0}} & & & 1 \\
\end{array}\right)
\begin{pmatrix}
1 & 0\\
0 & \mathcal{W}
\end{pmatrix},
\ee
such that
\begin{align}\label{eq:appR}
  R(\theta_i) =
  \left(\begin{array}{cc}
    \cos\theta_i & -\sin\theta_i \\
    \sin\theta_i & \cos\theta_i
  \end{array}\right),\quad (i = 1,...,N)
\end{align}
and $\mathcal{V}, \mathcal{W} \in SO(2N)$ satisfy
\begin{align}
  &\mathcal{V}_{2i} = \mathcal{V}_{2i-1} \Lambda,\quad (i = 1,...,N) \label{eq:appV}\\
  &\mathcal{W}_{2i} = \mathcal{W}_{2i-1} \Lambda,\quad (i = 1,...,N) \label{eq:appW}
\end{align}
where $\mathcal{V}_{n}$ ($\mathcal{W}_{n}$) stands for the $n$-th row of $\mathcal{V}$ ($\mathcal{W}$), and the $2N$$\times$$2N$ matrix $\Lambda$ is defined as
\begin{align}
  \Lambda =
  \mathbb{1}_N \otimes
  \begin{pmatrix}
    0 & 1 \\
    -1 & 0 \\
  \end{pmatrix} =
  \begin{pmatrix}
    0 & 1 & & & \\
    -1 & 0 & & \mbox{\LARGE{0}} & \\
    & & \ddots & & \\
    & \mbox{\LARGE{0}} & & 0 & 1 \\
    & & & -1 & 0 \\
  \end{pmatrix}. \label{eq:appLambda}
\end{align}

Before we start the proof, we point out that $\mathcal{V}$ ($\mathcal{W}$) satisfying Eq. \eqref{eq:appV} (Eq. \eqref{eq:appW}) is nothing but $\mathcal{V}$ ($\mathcal{W}$) given in Eq. \eqref{eq:vw_mf} written in a differently ordered basis (the purpose of such reordering is to make the central matrix at the right-hand-side of Eq. \eqref{eq:appdecO} taking a compact block-diagonal form). This can be shown by using explicitly
\begin{align}
  U_N
  = \frac{1}{\sqrt{2}}
  \begin{pmatrix}
    \mathbb{1}_N & \mathbb{1}_N \\
    i\mathbb{1}_N & -i\mathbb{1}_N \\
  \end{pmatrix}, \label{eq:appU}
\end{align}
hence we have, from Eq. \eqref{eq:vw_mf},
\begin{align}
  \mathcal{V} =
  \begin{pmatrix}
    \Re(V) & \Im(V) \\
    -\Im(V) & \Re(V) \\
  \end{pmatrix},\;
  \mathcal{W} =
  \begin{pmatrix}
    \Re(W) & \Im(W) \\
    -\Im(W) & \Re(W) \\
  \end{pmatrix}, \label{eq:appVW}
\end{align}
where $V$ and $W$ are $N$-dimensional unitary matrices (cf. Eqs. \eqref{eq:tf_ehv} and \eqref{eq:tf_ehw}). It is then straightforward to see that by rearranging, for $i = 1,...,N$, the $i$-th row and column to the ($2i-1$)-th and the ($i+N$)-th row and column to the ($2i$)-th, the rearranged $\mathcal{V}$ and $\mathcal{W}$ satisfy Eqs. \eqref{eq:appV} and \eqref{eq:appW}.

Now we prove our main claim by construction.

First we notice that $\mathcal{O}_N$ can always be written as
\be
  \mathcal{O}_N=
  \begin{pmatrix}
    \cos\theta_1 & (-\sin\theta_1)w\\
    (\sin\theta_1)v^T & X_0
  \end{pmatrix}
\ee
where both $v$ and $w$ are normalized $2N$-dimensional row vectors and $X_0$ is a $2N$$\times$$2N$ matrix. We assume $|\cos\theta_1|\ne 1$ (namely, the upper-left corner element of the matrix $\mathcal{O}_N$ is not $\pm1$), such that both $v$ and $w$ are well-defined up to an irrelevant sign.

Then we construct two matrices $\mathcal{V}^{(1)}, \mathcal{W}^{(1)} \in SO(2N)$ by setting
\begin{align}
  &\mathcal{V}_1^{(1)} = v,\;
  \mathcal{V}_2^{(1)} = \mathcal{V}_1^{(1)}\Lambda,\\
  &\mathcal{W}_1^{(1)} = w,\;
  \mathcal{W}_2^{(1)} = \mathcal{W}_1^{(1)}\Lambda,
\end{align}
and choosing $\mathcal{V}_i^{(1)}$ and $\mathcal{W}_i^{(1)}$ with $2< i \le 2N$ so that the orthogonality and Eqs. \eqref{eq:appV} and \eqref{eq:appW} are all satisfied (which can be easily shown to be always possible). We find
\begin{align}\label{eq:appStep1}
  &\begin{pmatrix}
    1 & 0\\
    0 & \mathcal{V}^{(1)}
  \end{pmatrix}
  \mathcal{O}_N
  \begin{pmatrix}
    1 & 0\\
    0 & {\mathcal{W}^{(1)}}^T
  \end{pmatrix} \nonumber\\
  =&\begin{pmatrix}
  \cos\theta_1 & -\sin\theta_1 & 0 & \ldots & 0 \\
  \sin\theta_1 & & & & \\
  0 & & & & \\
  \vdots & & \mbox{\huge{\textit{X}}}_1 & & \\
  0 & & & &
\end{pmatrix},
\end{align}
where $X_1 = \mathcal{V}^{(1)} X_0 {\mathcal{W}^{(1)}}^T$.

Since all three factor matrices at the left-hand-side of the above equation \eqref{eq:appStep1} belong to $SO(2N+1)$, so must do their product. It follows that
\be
\label{eq:appX1}
X_1=\begin{pmatrix}
\cos\theta_1 & 0\\
0 & \mathcal{O}_{N-1}
\end{pmatrix},
\ee
where $\mathcal{O}_{N-1} \in SO(2N-1)$. In other words, we have
\begin{align}\label{eq:appdec1}
  \begin{pmatrix}
    1 & 0\\
    0 & \mathcal{V}^{(1)}
  \end{pmatrix}
  \mathcal{O}_N
  \begin{pmatrix}
    1 & 0\\
    0 & {\mathcal{W}^{(1)}}^T
  \end{pmatrix}
  =
  \begin{pmatrix}
  R(\theta_1) & 0 \\
  0 & \mathcal{O}_{N-1}
  \end{pmatrix}.
\end{align}

Obviously this procedure can be continued for $\mathcal{O}_{N-1}$, and then for $\mathcal{O}_{N-2}$ etc., until we arrive at $\mathcal{O}_{0}$, which is $1$. Meanwhile we obtain Eq. \eqref{eq:appdecO} with
\begin{align}
  &\mathcal{V} =
  \begin{pmatrix}
    \mathbb{1}_{2(N-1)} & 0\\
    0 & \mathcal{V}^{(N)}
  \end{pmatrix}
  \cdots
  \begin{pmatrix}
    \mathbb{1}_{2} & 0\\
    0 & \mathcal{V}^{(2)}
  \end{pmatrix}
  \mathcal{V}^{(1)}, \label{eq:appVc}\\
  &\mathcal{W} =
  \begin{pmatrix}
    \mathbb{1}_{2(N-1)} & 0\\
    0 & \mathcal{W}^{(N)}
  \end{pmatrix}
  \cdots
  \begin{pmatrix}
    \mathbb{1}_{2} & 0\\
    0 & \mathcal{W}^{(2)}
  \end{pmatrix}
  \mathcal{W}^{(1)}. \label{eq:appWc}
\end{align}

It remains to show that $\mathcal{V}$ and $\mathcal{W}$ defined above satisfy Eqs. \eqref{eq:appV} and \eqref{eq:appW}. To this end we observe $\Lambda^2 = -\mathbb{1}$, hence Eqs. \eqref{eq:appV} and \eqref{eq:appW} are equivalent to
\begin{align}
  [\mathcal{V}, \Lambda] = 0,\quad [\mathcal{W}, \Lambda] = 0,
\end{align}
where $[,]$ stands for the commutator. We also keep in mind that, by construction, all factor matrices at the right-hand-sides of Eqs. \eqref{eq:appVc} and \eqref{eq:appWc} satisfy the above commutation relations. This immediately verifies the validity of our construction, and thus completes the proof.


\begin{thebibliography}{36}
\expandafter\ifx\csname natexlab\endcsname\relax\def\natexlab#1{#1}\fi
\expandafter\ifx\csname bibnamefont\endcsname\relax
  \def\bibnamefont#1{#1}\fi
\expandafter\ifx\csname bibfnamefont\endcsname\relax
  \def\bibfnamefont#1{#1}\fi
\expandafter\ifx\csname citenamefont\endcsname\relax
  \def\citenamefont#1{#1}\fi
\expandafter\ifx\csname url\endcsname\relax
  \def\url#1{\texttt{#1}}\fi
\expandafter\ifx\csname urlprefix\endcsname\relax\def\urlprefix{URL }\fi
\providecommand{\bibinfo}[2]{#2}
\providecommand{\eprint}[2][]{\url{#2}}

\bibitem[{\citenamefont{Kitaev}(2001)}]{kitaev_unpaired_2001}
\bibinfo{author}{\bibfnamefont{A.~Y.} \bibnamefont{Kitaev}},
  \bibinfo{journal}{{Physics-Uspekhi}} \textbf{\bibinfo{volume}{44}},
  \bibinfo{pages}{131} (\bibinfo{year}{2001}).

\bibitem[{\citenamefont{Read and Green}(2000)}]{read_paired_2000}
\bibinfo{author}{\bibfnamefont{N.}~\bibnamefont{Read}} \bibnamefont{and}
  \bibinfo{author}{\bibfnamefont{D.}~\bibnamefont{Green}},
  \bibinfo{journal}{Physical Review B} \textbf{\bibinfo{volume}{61}},
  \bibinfo{pages}{10267} (\bibinfo{year}{2000}).

\bibitem[{\citenamefont{Schnyder et~al.}(2008)\citenamefont{Schnyder, Ryu,
  Furusaki, and Ludwig}}]{schnyder_classification_2008}
\bibinfo{author}{\bibfnamefont{A.~P.} \bibnamefont{Schnyder}},
  \bibinfo{author}{\bibfnamefont{S.}~\bibnamefont{Ryu}},
  \bibinfo{author}{\bibfnamefont{A.}~\bibnamefont{Furusaki}}, \bibnamefont{and}
  \bibinfo{author}{\bibfnamefont{A.~W.~W.} \bibnamefont{Ludwig}},
  \bibinfo{journal}{Physical Review B} \textbf{\bibinfo{volume}{78}},
  \bibinfo{pages}{195125} (\bibinfo{year}{2008}).

\bibitem[{\citenamefont{Schnyder et~al.}(2009)\citenamefont{Schnyder, Ryu,
  Furusaki, and Ludwig}}]{schnyder_classification_2009}
\bibinfo{author}{\bibfnamefont{A.~P.} \bibnamefont{Schnyder}},
  \bibinfo{author}{\bibfnamefont{S.}~\bibnamefont{Ryu}},
  \bibinfo{author}{\bibfnamefont{A.}~\bibnamefont{Furusaki}}, \bibnamefont{and}
  \bibinfo{author}{\bibfnamefont{A.~W.~W.} \bibnamefont{Ludwig}},
  \bibinfo{journal}{{AIP} Conference Proceedings}
  \textbf{\bibinfo{volume}{1134}}, \bibinfo{pages}{10} (\bibinfo{year}{2009}).

\bibitem[{\citenamefont{Qi and Zhang}(2011)}]{qi_topological_2011}
\bibinfo{author}{\bibfnamefont{X.}~\bibnamefont{Qi}} \bibnamefont{and}
  \bibinfo{author}{\bibfnamefont{S.}~\bibnamefont{Zhang}},
  \bibinfo{journal}{Reviews of Modern Physics} \textbf{\bibinfo{volume}{83}},
  \bibinfo{pages}{1057} (\bibinfo{year}{2011}).

\bibitem[{\citenamefont{Fu and Kane}(2009{\natexlab{a}})}]{fu_josephson_2009}
\bibinfo{author}{\bibfnamefont{L.}~\bibnamefont{Fu}} \bibnamefont{and}
  \bibinfo{author}{\bibfnamefont{C.~L.} \bibnamefont{Kane}},
  \bibinfo{journal}{Physical Review B} \textbf{\bibinfo{volume}{79}},
  \bibinfo{pages}{161408} (\bibinfo{year}{2009}{\natexlab{a}}).

\bibitem[{\citenamefont{Nilsson et~al.}(2008)\citenamefont{Nilsson, Akhmerov,
  and Beenakker}}]{nilsson_splitting_2008}
\bibinfo{author}{\bibfnamefont{J.}~\bibnamefont{Nilsson}},
  \bibinfo{author}{\bibfnamefont{A.~R.} \bibnamefont{Akhmerov}},
  \bibnamefont{and} \bibinfo{author}{\bibfnamefont{C.~W.~J.}
  \bibnamefont{Beenakker}}, \bibinfo{journal}{Physical Review Letters}
  \textbf{\bibinfo{volume}{101}}, \bibinfo{pages}{120403}
  (\bibinfo{year}{2008}).

\bibitem[{\citenamefont{Sau et~al.}(2010)\citenamefont{Sau, Lutchyn, Tewari,
  and Das~Sarma}}]{sau_generic_2010}
\bibinfo{author}{\bibfnamefont{J.~D.} \bibnamefont{Sau}},
  \bibinfo{author}{\bibfnamefont{R.~M.} \bibnamefont{Lutchyn}},
  \bibinfo{author}{\bibfnamefont{S.}~\bibnamefont{Tewari}}, \bibnamefont{and}
  \bibinfo{author}{\bibfnamefont{S.}~\bibnamefont{Das~Sarma}},
  \bibinfo{journal}{Physical Review Letters} \textbf{\bibinfo{volume}{104}},
  \bibinfo{pages}{040502} (\bibinfo{year}{2010}).

\bibitem[{\citenamefont{Lutchyn et~al.}(2010)\citenamefont{Lutchyn, Sau, and
  Das~Sarma}}]{lutchyn_majorana_2010}
\bibinfo{author}{\bibfnamefont{R.~M.} \bibnamefont{Lutchyn}},
  \bibinfo{author}{\bibfnamefont{J.~D.} \bibnamefont{Sau}}, \bibnamefont{and}
  \bibinfo{author}{\bibfnamefont{S.}~\bibnamefont{Das~Sarma}},
  \bibinfo{journal}{Physical Review Letters} \textbf{\bibinfo{volume}{105}},
  \bibinfo{pages}{077001} (\bibinfo{year}{2010}).

\bibitem[{\citenamefont{Oreg et~al.}(2010)\citenamefont{Oreg, Refael, and von
  Oppen}}]{oreg_helical_2010}
\bibinfo{author}{\bibfnamefont{Y.}~\bibnamefont{Oreg}},
  \bibinfo{author}{\bibfnamefont{G.}~\bibnamefont{Refael}}, \bibnamefont{and}
  \bibinfo{author}{\bibfnamefont{F.}~\bibnamefont{von Oppen}},
  \bibinfo{journal}{Physical Review Letters} \textbf{\bibinfo{volume}{105}},
  \bibinfo{pages}{177002} (\bibinfo{year}{2010}).

\bibitem[{\citenamefont{Akhmerov et~al.}(2011)\citenamefont{Akhmerov, Dahlhaus,
  Hassler, Wimmer, and Beenakker}}]{akhmerov_quantized_2011}
\bibinfo{author}{\bibfnamefont{A.~R.} \bibnamefont{Akhmerov}},
  \bibinfo{author}{\bibfnamefont{J.~P.} \bibnamefont{Dahlhaus}},
  \bibinfo{author}{\bibfnamefont{F.}~\bibnamefont{Hassler}},
  \bibinfo{author}{\bibfnamefont{M.}~\bibnamefont{Wimmer}}, \bibnamefont{and}
  \bibinfo{author}{\bibfnamefont{C.~W.~J.} \bibnamefont{Beenakker}},
  \bibinfo{journal}{Physical Review Letters} \textbf{\bibinfo{volume}{106}},
  \bibinfo{pages}{057001} (\bibinfo{year}{2011}).

\bibitem[{\citenamefont{Wimmer et~al.}(2011)\citenamefont{Wimmer, Akhmerov,
  Dahlhaus, and Beenakker}}]{wimmer_quantum_2011}
\bibinfo{author}{\bibfnamefont{M.}~\bibnamefont{Wimmer}},
  \bibinfo{author}{\bibfnamefont{A.~R.} \bibnamefont{Akhmerov}},
  \bibinfo{author}{\bibfnamefont{J.~P.} \bibnamefont{Dahlhaus}},
  \bibnamefont{and} \bibinfo{author}{\bibfnamefont{C.~W.~J.}
  \bibnamefont{Beenakker}}, \bibinfo{journal}{New Journal of Physics}
  \textbf{\bibinfo{volume}{13}}, \bibinfo{pages}{053016}
  (\bibinfo{year}{2011}).

\bibitem[{\citenamefont{Alicea et~al.}(2011)\citenamefont{Alicea, Oreg, Refael,
  von Oppen, and Fisher}}]{alicea_non-abelian_2011}
\bibinfo{author}{\bibfnamefont{J.}~\bibnamefont{Alicea}},
  \bibinfo{author}{\bibfnamefont{Y.}~\bibnamefont{Oreg}},
  \bibinfo{author}{\bibfnamefont{G.}~\bibnamefont{Refael}},
  \bibinfo{author}{\bibfnamefont{F.}~\bibnamefont{von Oppen}},
  \bibnamefont{and} \bibinfo{author}{\bibfnamefont{M.~P.~A.}
  \bibnamefont{Fisher}}, \bibinfo{journal}{Nat Phys}
  \textbf{\bibinfo{volume}{7}}, \bibinfo{pages}{412} (\bibinfo{year}{2011}).

\bibitem[{\citenamefont{Flensberg}(2011)}]{flensberg_non-abelian_2011}
\bibinfo{author}{\bibfnamefont{K.}~\bibnamefont{Flensberg}},
  \bibinfo{journal}{Physical Review Letters} \textbf{\bibinfo{volume}{106}},
  \bibinfo{pages}{090503} (\bibinfo{year}{2011}).

\bibitem[{\citenamefont{Beenakker}(2011)}]{beenakker_review_2012}
\bibinfo{author}{\bibfnamefont{C.~W.~J.} \bibnamefont{Beenakker}},
  \bibinfo{journal}{{arXiv:1112.1950}}  (\bibinfo{year}{2011}).

\bibitem[{\citenamefont{Fu and Kane}(2008)}]{fu_superconducting_2008}
\bibinfo{author}{\bibfnamefont{L.}~\bibnamefont{Fu}} \bibnamefont{and}
  \bibinfo{author}{\bibfnamefont{C.~L.} \bibnamefont{Kane}},
  \bibinfo{journal}{Physical Review Letters} \textbf{\bibinfo{volume}{100}},
  \bibinfo{pages}{096407} (\bibinfo{year}{2008}).

\bibitem[{\citenamefont{Fu and Kane}(2009{\natexlab{b}})}]{fu_probing_2009}
\bibinfo{author}{\bibfnamefont{L.}~\bibnamefont{Fu}} \bibnamefont{and}
  \bibinfo{author}{\bibfnamefont{C.~L.} \bibnamefont{Kane}},
  \bibinfo{journal}{Physical Review Letters} \textbf{\bibinfo{volume}{102}},
  \bibinfo{pages}{216403} (\bibinfo{year}{2009}{\natexlab{b}}).

\bibitem[{\citenamefont{Akhmerov et~al.}(2009)\citenamefont{Akhmerov, Nilsson,
  and Beenakker}}]{akhmerov_electrically_2009}
\bibinfo{author}{\bibfnamefont{A.~R.} \bibnamefont{Akhmerov}},
  \bibinfo{author}{\bibfnamefont{J.}~\bibnamefont{Nilsson}}, \bibnamefont{and}
  \bibinfo{author}{\bibfnamefont{C.~W.~J.} \bibnamefont{Beenakker}},
  \bibinfo{journal}{Physical Review Letters} \textbf{\bibinfo{volume}{102}},
  \bibinfo{pages}{216404} (\bibinfo{year}{2009}).

\bibitem[{\citenamefont{Law et~al.}(2009)\citenamefont{Law, Lee, and
  Ng}}]{law_majorana_2009}
\bibinfo{author}{\bibfnamefont{K.~T.} \bibnamefont{Law}},
  \bibinfo{author}{\bibfnamefont{P.~A.} \bibnamefont{Lee}}, \bibnamefont{and}
  \bibinfo{author}{\bibfnamefont{T.~K.} \bibnamefont{Ng}},
  \bibinfo{journal}{Physical Review Letters} \textbf{\bibinfo{volume}{103}},
  \bibinfo{pages}{237001} (\bibinfo{year}{2009}).

\bibitem[{\citenamefont{Str\"{u}bi et~al.}(2011)\citenamefont{Str\"{u}bi,
  Belzig, Choi, and Bruder}}]{struebi_interferometric_2011}
\bibinfo{author}{\bibfnamefont{G.}~\bibnamefont{Str\"{u}bi}},
  \bibinfo{author}{\bibfnamefont{W.}~\bibnamefont{Belzig}},
  \bibinfo{author}{\bibfnamefont{M.}~\bibnamefont{Choi}}, \bibnamefont{and}
  \bibinfo{author}{\bibfnamefont{C.}~\bibnamefont{Bruder}},
  \bibinfo{journal}{Physical Review Letters} \textbf{\bibinfo{volume}{107}},
  \bibinfo{pages}{136403} (\bibinfo{year}{2011}).

\bibitem[{\citenamefont{Kopnin and Salomaa}(1991)}]{kopnin_mutual_1991}
\bibinfo{author}{\bibfnamefont{N.~B.} \bibnamefont{Kopnin}} \bibnamefont{and}
  \bibinfo{author}{\bibfnamefont{M.~M.} \bibnamefont{Salomaa}},
  \bibinfo{journal}{Physical Review B} \textbf{\bibinfo{volume}{44}},
  \bibinfo{pages}{9667} (\bibinfo{year}{1991}).

\bibitem[{\citenamefont{Ivanov}(2001)}]{ivanov_non-abelian_2001}
\bibinfo{author}{\bibfnamefont{D.~A.} \bibnamefont{Ivanov}},
  \bibinfo{journal}{Physical Review Letters} \textbf{\bibinfo{volume}{86}},
  \bibinfo{pages}{268} (\bibinfo{year}{2001}).

\bibitem[{\citenamefont{Altland and
  Zirnbauer}(1997)}]{altland_nonstandard_1997}
\bibinfo{author}{\bibfnamefont{A.}~\bibnamefont{Altland}} \bibnamefont{and}
  \bibinfo{author}{\bibfnamefont{M.~R.} \bibnamefont{Zirnbauer}},
  \bibinfo{journal}{Physical Review B} \textbf{\bibinfo{volume}{55}},
  \bibinfo{pages}{1142} (\bibinfo{year}{1997}).

\bibitem[{\citenamefont{Landauer}(1970)}]{landauer_electrical_1970}
\bibinfo{author}{\bibfnamefont{R.}~\bibnamefont{Landauer}},
  \bibinfo{journal}{Philosophical Magazine} \textbf{\bibinfo{volume}{21}},
  \bibinfo{pages}{863} (\bibinfo{year}{1970}).

\bibitem[{\citenamefont{B\"{u}ttiker}(1986)}]{buettiker_four-terminal_1986}
\bibinfo{author}{\bibfnamefont{M.}~\bibnamefont{B\"{u}ttiker}},
  \bibinfo{journal}{Physical Review Letters} \textbf{\bibinfo{volume}{57}},
  \bibinfo{pages}{1761} (\bibinfo{year}{1986}).

\bibitem[{\citenamefont{B\"{u}ttiker}(1992)}]{buttiker_scattering_1992}
\bibinfo{author}{\bibfnamefont{M.}~\bibnamefont{B\"{u}ttiker}},
  \bibinfo{journal}{Physical Review B} \textbf{\bibinfo{volume}{46}},
  \bibinfo{pages}{12485} (\bibinfo{year}{1992}).

\bibitem[{\citenamefont{Fulga et~al.}(2011{\natexlab{a}})\citenamefont{Fulga,
  Hassler, Akhmerov, and Beenakker}}]{fulga_scattering_formula_2011}
\bibinfo{author}{\bibfnamefont{I.~C.} \bibnamefont{Fulga}},
  \bibinfo{author}{\bibfnamefont{F.}~\bibnamefont{Hassler}},
  \bibinfo{author}{\bibfnamefont{A.~R.} \bibnamefont{Akhmerov}},
  \bibnamefont{and} \bibinfo{author}{\bibfnamefont{C.~W.~J.}
  \bibnamefont{Beenakker}}, \bibinfo{journal}{Physical Review B}
  \textbf{\bibinfo{volume}{83}}, \bibinfo{pages}{155429}
  (\bibinfo{year}{2011}{\natexlab{a}}).

\bibitem[{\citenamefont{Meidan et~al.}(2011)\citenamefont{Meidan, Micklitz, and
  Brouwer}}]{meidan_topological_2011}
\bibinfo{author}{\bibfnamefont{D.}~\bibnamefont{Meidan}},
  \bibinfo{author}{\bibfnamefont{T.}~\bibnamefont{Micklitz}}, \bibnamefont{and}
  \bibinfo{author}{\bibfnamefont{P.~W.} \bibnamefont{Brouwer}},
  \bibinfo{journal}{Physical Review B} \textbf{\bibinfo{volume}{84}},
  \bibinfo{pages}{195410} (\bibinfo{year}{2011}).

\bibitem[{\citenamefont{Fulga et~al.}(2011{\natexlab{b}})\citenamefont{Fulga,
  Hassler, and Akhmerov}}]{fulga_scattering_theory_2011}
\bibinfo{author}{\bibfnamefont{I.~C.} \bibnamefont{Fulga}},
  \bibinfo{author}{\bibfnamefont{F.}~\bibnamefont{Hassler}}, \bibnamefont{and}
  \bibinfo{author}{\bibfnamefont{A.~R.} \bibnamefont{Akhmerov}},
  \bibinfo{journal}{{arXiv:1106.6351}}  (\bibinfo{year}{2011}{\natexlab{b}}).

\bibitem[{\citenamefont{Graf and Ortelli}(2011)}]{graf_topological_2008}
\bibinfo{author}{\bibfnamefont{G.~M.} \bibnamefont{Graf}} \bibnamefont{and}
  \bibinfo{author}{\bibfnamefont{G.}~\bibnamefont{Ortelli}},
  \bibinfo{journal}{Physical Review B} \textbf{\bibinfo{volume}{77}},
  \bibinfo{pages}{195410} (\bibinfo{year}{2011}).

\bibitem[{\citenamefont{Anantram and Datta}(1996)}]{anantram_current_1996}
\bibinfo{author}{\bibfnamefont{M.~P.} \bibnamefont{Anantram}} \bibnamefont{and}
  \bibinfo{author}{\bibfnamefont{S.}~\bibnamefont{Datta}},
  \bibinfo{journal}{Physical Review B} \textbf{\bibinfo{volume}{53}},
  \bibinfo{pages}{16390} (\bibinfo{year}{1996}).

\bibitem[{\citenamefont{Mello et~al.}(1988)\citenamefont{Mello, Pereyra, and
  Kumar}}]{mello_macroscopic_1988}
\bibinfo{author}{\bibfnamefont{P.~A.} \bibnamefont{Mello}},
  \bibinfo{author}{\bibfnamefont{P.}~\bibnamefont{Pereyra}}, \bibnamefont{and}
  \bibinfo{author}{\bibfnamefont{N.}~\bibnamefont{Kumar}},
  \bibinfo{journal}{Annals of Physics} \textbf{\bibinfo{volume}{181}},
  \bibinfo{pages}{290} (\bibinfo{year}{1988}).

\bibitem[{\citenamefont{Martin and Landauer}(1992)}]{martin_wave-packet_1992}
\bibinfo{author}{\bibfnamefont{T.}~\bibnamefont{Martin}} \bibnamefont{and}
  \bibinfo{author}{\bibfnamefont{R.}~\bibnamefont{Landauer}},
  \bibinfo{journal}{Physical Review B} \textbf{\bibinfo{volume}{45}},
  \bibinfo{pages}{1742} (\bibinfo{year}{1992}).

\bibitem[{\citenamefont{Pikulin and Nazarov}(2011)}]{pikulin_topological_2011}
\bibinfo{author}{\bibfnamefont{D.~I.} \bibnamefont{Pikulin}} \bibnamefont{and}
  \bibinfo{author}{\bibfnamefont{Y.~V.} \bibnamefont{Nazarov}},
  \bibinfo{journal}{{arXiv:1103.0780}}  (\bibinfo{year}{2011}).

\bibitem[{\citenamefont{Serban et~al.}(2010)\citenamefont{Serban, B\'{e}ri,
  Akhmerov, and Beenakker}}]{serban_cre_2010}
\bibinfo{author}{\bibfnamefont{I.}~\bibnamefont{Serban}},
  \bibinfo{author}{\bibfnamefont{B.}~\bibnamefont{B\'{e}ri}},
  \bibinfo{author}{\bibfnamefont{A.~R.} \bibnamefont{Akhmerov}},
  \bibnamefont{and} \bibinfo{author}{\bibfnamefont{C.~W.~J.}
  \bibnamefont{Beenakker}}, \bibinfo{journal}{Physical Review Letters}
  \textbf{\bibinfo{volume}{104}}, \bibinfo{pages}{147001}
  (\bibinfo{year}{2010}).

\bibitem[{\citenamefont{B\'{e}ri}(2009)}]{beri_dephasing-enabled_2009}
\bibinfo{author}{\bibfnamefont{B.}~\bibnamefont{B\'{e}ri}},
  \bibinfo{journal}{Physical Review B} \textbf{\bibinfo{volume}{79}},
  \bibinfo{pages}{245315} (\bibinfo{year}{2009}).

\end{thebibliography}

\end{document}